\documentclass[
superscriptaddress,
preprint,
noshowpacs,
 amsmath,amssymb,
 aps,
]{revtex4-1}

\usepackage{graphicx}
\usepackage{dcolumn}
\usepackage{bm}

\usepackage{url}
\begin{document}


\title{\textbf{Ultrasensitive and highly accurate long-range surface plasmon resonance biosensors based on two-dimensional transition metal dichalcogenides}}

\author{Yi Xu}
\affiliation{%
 Engineering Product Development, \\Singapore University of Technology and Design, Singapore 487372, Singapore
}%
\author{Chang-Yu Hsieh}
\affiliation{%
 Department of Chemistry, Massachusetts Institute of Technology, 77 Massachusetts Avenue, Cambridge, MA 02139, USA
}%
\affiliation{Singapore-MIT Alliance for Research and Technology (SMART) center, 1 CREATE Way, Singapore 138602, Singapore}

\author{Lin Wu}%
\affiliation{%
Electronics and Photonics Department, Institute of High Performance Computing, \\ Agency for Science, Technology, and Research (A*STAR), Singapore 138632, Singapore}%

\author{L. K. Ang}%
 \email{ricky\_ang@sutd.edu.sg}
\affiliation{%
 Engineering Product Development, \\Singapore University of Technology and Design, Singapore 487372, Singapore
}%

\date{\today}

\begin{abstract}
Two-dimensional transition metal dichalcogenides (TMDCs), as promising alternative plasmonics supporting materials to graphene, exhibit potential applications in sensing. Here, we propose an ultrasensitive, accurate long-range surface plasmon resonance (LRSPR) imaging biosensor with two-dimensional TMDC layers, which shows higher detection accuracy than that of conventional SPR biosensor. It is found that the imaging sensitivity of the proposed LRSPR biosensor can be enhanced by the integration of TMDC layers, which is different from the previous graphene-based LRSPR or SPR imaging sensor, whose imaging sensitivity usually decreases with the number of graphene layers. The sensitivity enhancement or degradation effect for the proposed chalcogenide-cytop-gold-TMDCs based biosensor depends on the thickness of gold thin film and cytop layer. Imaging sensitivity of more than 4000 $\text{RIU}^{-1}$ can be obtained with a high detection accuracy of more than 120 $\text{deg}^{-1}$. We expect that the proposed TMDCs mediated LRSPR imaging sensor could provide potential applications in chemical sensing and biosensing for a highly sensitive and accurate simultaneous detection of multiple biomolecular interactions.

\end{abstract}

\maketitle


\section{Introduction}
Surface plasmon resonance (SPR) has been widely employed for sensing applications \cite{homola1999surface,wijaya2011surface,shalabney2011sensitivity,homola2008surface}, such as gas sensing, temperature sensing, and biosensing, during the last two decades due to its high sensitivity and reliability. In general, the sensing principle of SPR sensor is the utilization of the exponentially decaying fields of a surface plasmon wave (SPW) propagating along the metal-dielectric interface, which is highly sensitive to the ambient refractive index (RI) variations, such as induced by bioaffinity interactions at the sensor surface. One of the common techniques for SPWs excitation is the Kretschmann configurtion \cite{kretschmann1968notizen}, in which the base of glass prism is coated with a metal thin film, and the SPW  at the metal-sening layer interface was excited by a p-polarized incident light beam when the parallel component of the incident light wave vector $k_x$ matches with the wave vector of SPW $k_{sp}$:
\begin{equation}
\frac{2 \pi}{\lambda_0}n_{prism}\sin \theta = k_x = \text{Re}\{k_{sp}\}=\text{Re}\left\{\frac{2 \pi}{\lambda_0}\sqrt{\frac{\varepsilon_m \varepsilon_d}{\varepsilon_m+\varepsilon_d}}\right\},
\end{equation}
where $\lambda_0$ is the incident wavelength in vacuum, $n_{prism}$ is the RI of prism, $\theta$ is the incident angle, $\varepsilon_m$ and $\varepsilon_d$ are the dielectric constants of the metal film and dielectric layer (i.e., sensing layer), respectively. The successful excitation of SPWs results in a minimum reflectance whose value and position are extremely sensitive to the sensing layer RI variations. For SPR sensor with angle modulation, the wavelength of incident light is fixed, and the resonance angle serves as an output signal of the SPR sensor. However, one disadvantage with the angular interrogation technique is that it does not allow the parallel monitoring of numerous biomolecular interactions at a time. SPR imaging sensors have been proposed and demonstrated to overcome the limits on the parallel monitoring \cite{lee2006surface,choi2010investigation,choi2011graphene,wong2014surface,zeng2017recent,wark2005long}. In the SPR imaging sensor, the incident angle is fixed, the spatial variations in reflectivity induced by the ambient RI changes are measured. In addition, the SPR imaging sensor is more manoeuvrable since the imaging technique does not require movement of any components of the SPR imaging sensor. 

Another problem of the SPR sensor is its broad SPR curve, which limits the detection accuracy (DA). Long-rang SPR (LRSPR) is an effective way to improve the sensor's DA and sensitivity. Long-rang surface plasmons (LRSPs) first predicted by Sarid \cite{sarid1981long} are surface electromagnetic waves propagating along thin metal film that embedded between two dielectric layers with similar RIs. LRSPR has narrower reflectance-angle curves, longer evanescent field penetration depth and higher electric field at the metal-dielectric interface as compared to conventional SPR (cSPR) \cite{dostalek2007long,slavik2007ultrahigh,nenninger2001long,chabot2012long,wark2005long,berini2009long}, which in turn improves the sensitivity and DA of SPR sensors. 

Graphene, a two-dimensional (2D) sheet of carbon atoms, has been extensively studied in recent years due to its fascinating physical and chemical properties \cite{geim2007rise,novoselov2012roadmap,ferrari2015science}. As a plasmonics supporting material \cite{koppens2011graphene,grigorenko2012graphene}, graphene has been employed for sensing applications \cite{kim2013graphene,szunerits2013recent}. In addition, the combination of graphene and metal thin film has demonstrated to improve the sensor sensitivity \cite{wu2010highly,fu2015graphene,mishra2016graphene}. For example, an angular sensitivity enhancement of 25\% can be achieved with 10 layers of graphene deposited on gold (Au) film for prism-coupled SPR sensor \cite{wu2010highly}. The successful applications of graphene in sensing area has ignited the motivation to explore other 2D material in sensing applications. One example is the 2D transition metal dichalcogenides (TMDCs), which consist of transition metal atom M (like Mo, W) and chalcogen atom X (such as S, Se) with a general chemical formula $\text{MX}_2$. The exciting optical, electrical, chemical properties of TMDCs \cite{jariwala2014emerging,wang2012electronics,mak2016photonics,duan2015two} making them as promising candidates for future nanoelectronic and optoelectronic applications. TMDCs-based SPR sensors have been proposed for RI sensing \cite{zeng2015agraphene,ouyang2016sensitivity,ouyang2017two,mishra2016graphene}, which exhibit enhanced sensitivity. Thus, a question arises: which material system has a better sensor performance, graphene or TMDCs? 

Taking the advantages of SPR imaging sensor, LRSPR and TMDCs, we propose an ultrasensitive and high accurate LRSPR imaging biosensor with four TMDC materials: Molybdenum disulfide ($\text{MoS}_2$), Molybdenum diselenide ($\text{MoSe}_2$), Tungsten disulfide ($\text{WS}_2$), and Tungsten diselenide ($\text{WSe}_2$). In the proposed sensor configurations (chalcogenide(2S2G)-cytop-Au-TMDCs), TMDC layers, directly contacted with biomolecules (analyte), act as a signal-enhanced layer due to a high efficiency of charge transfer between TMDC layer and Au surface \cite{ouyang2016sensitivity,hoggard2013using,fang2013gated,giovannetti2008doping}. In addition, the TMDC layers also serve as biomolecules absorption mediums \cite{lee2014two,zhu2013single,farimani2014dna}. The imaging sensitivity enhancement or degradation effect for the proposed LRSPR biosensor are investigated, which is found to be dependent on the thickness of Au thin film and cytop layer. The proposed TMDCs mediated LRSPR imaging sensor shows an ultrahigh imaging sensitivity of more than 4000 $\text{RIU}^{-1}$ with a high DA of more than 120 $\text{deg}^{-1}$.
\section{Design consideration and theoretical model}
\subsection{RIs of various layers}
In the designed SPR sensor, the coupling prism is 2S2G glass prism, a promising candidate for the design of SPR sensor due to its high RI and broad operating window. The wavelength-dependent RI of the 2S2G prism is \cite{maharana2012chalcogenide}
\begin{equation}
n_{\text{2S2G}}=2.24047+\frac{2.693\times10^{-2}}{\lambda^2} + \frac{9.08\times10^{-3}}{\lambda^4},
\end{equation}
where the wavelength $\lambda$ is given in $\mu$m. The RI of Au film is given by 
\begin{equation}
n_{\text{Au}}=\left(1-\frac{\lambda^2\lambda_c}{\lambda^{2}_p(\lambda_c+i\lambda)}\right)^{1/2},
\end{equation}
according to the Drude-Lorentz model \cite{maharana2014performance}. Here,    $\lambda_p$ ($1.6826\times10^{-7}$ m) and $\lambda_c$ ($8.9342\times10^{-6}$ m) are the plasma wavelength and collision wavelength of Au, respectively. The thickness of monolayer graphene is 0.34 nm and its RI in the visible range is given by \cite{bruna2009optical} 
\begin{equation}
n_{\text{graphene}}=3.0+i \frac{C_1}{3}\lambda,
\end{equation}
where the constant $C_1\approx 5.446\mu\text{m}^{-1}$, and $\lambda$ is the wavelength in $\mu\text{m}$. The thickness of four TMDC materials, $\text{MoS}_2$, $\text{MoSe}_2$, $\text{WS}_2$, $\text{WSe}_2$, are 0.65 nm, 0.70 nm, 0.80 and 0.70 nm respectively, with the complex RIs of $5.0805+i 1.1723$, $4.6226+i 1.0063$, $4.8937+ i 0.3124$, $4.5501+i 0.4332$ at wavelength $\lambda=633$ nm \cite{ouyang2017two}. The RI of cytop layer in LRSPR sensor is 1.3395 at $\lambda=633$ nm \cite{cytop}, which is close to the RI of sensing layer, $n_s=1.330$. 

\subsection{Sensor performance parameter: sensitivity and detection accuracy}
The reflectance of a prism-coupled SPR sensor can be calculate with a generalized N-layer model \cite{yamamoto2002surface}. For p-polarized incident light beam, the reflectance $R_p$ is given by 
\begin{equation}
R_p=\left| \frac{(M_{11}+M_{12}q_N)q_1 - (M_{21}+M_{22}q_N)}{(M_{11}+M_{12}q_N)q_1 + (M_{21}+M_{22}q_N)}\right|^2,
\label{eq:R}
\end{equation}
with 
\begin{equation}
M=\prod_{k=2}^{N-1}M_k=\left[\begin{array}{cc}
M_{11} & M_{12} \\
M_{21} & M_{22}
\end{array}\right],
\end{equation}
where
\begin{equation}
M_k=\left[\begin{array}{cc}
\cos \beta_k & -i (\sin \beta_k)/q_k \\
-i q_k \sin \beta_k & \cos \beta_k
\end{array}\right],
\end{equation}
 
\begin{equation}
\beta_k=\frac{2\pi d_k}{\lambda}\left(n_k^2-n_1^2\sin^2\theta_1\right),
\end{equation}
and 
\begin{equation}
q_k=\frac{\left(n_k^2-n_1^2\sin^2\theta_1\right)^{1/2}}{n_k^2},
\end{equation}
Here, $n_k$ and $d_k$ are respectively the RI and thickness of the $k$th layer with $k$ varies from 2 to $N-1$. The first layer and the $N$th layer are the 2S2G prism and sensing layer, respectively. $\lambda$ is the wavelength of excitation light, and $\theta_1$ is the incident angle. A variation of the sensing layer RI will cause a change in the resonance angle $\theta_{res}$ as well as the reflectance $R_p$, then the angular sensitivity of SPR sensor is defined as 
\begin{equation}
S_{angle}=\frac{\Delta \theta_{res}}{\Delta n_s}.
\end{equation}
For SPR imaging sensor, the spatial changes in reflectance $R_p$, induced by the ambient RI variations, are measured at a fixed incident angle. The imaging sensitivity is given by
\begin{equation}
S_{imaging}=\frac{d R_p}{d n_s}.
\end{equation} 
In addition to the sensitivity, another important sensor performance parameter is DA, which is defined as the reciprocal of full width at half maximum (FWHM):
\begin{equation}
DA=\frac{1}{\text{FWHM}}.
\end{equation} 
Narrower FWHM, i.e., higher DA, helps the accurate measurement of the reflectance minimum or resonance angle. Therefore, for a SPR sensor with excellent performance, both the sensitivity and the DA should be as high as possible. 
\begin{figure}[thpb]
      \centering
      \includegraphics[scale=0.4]{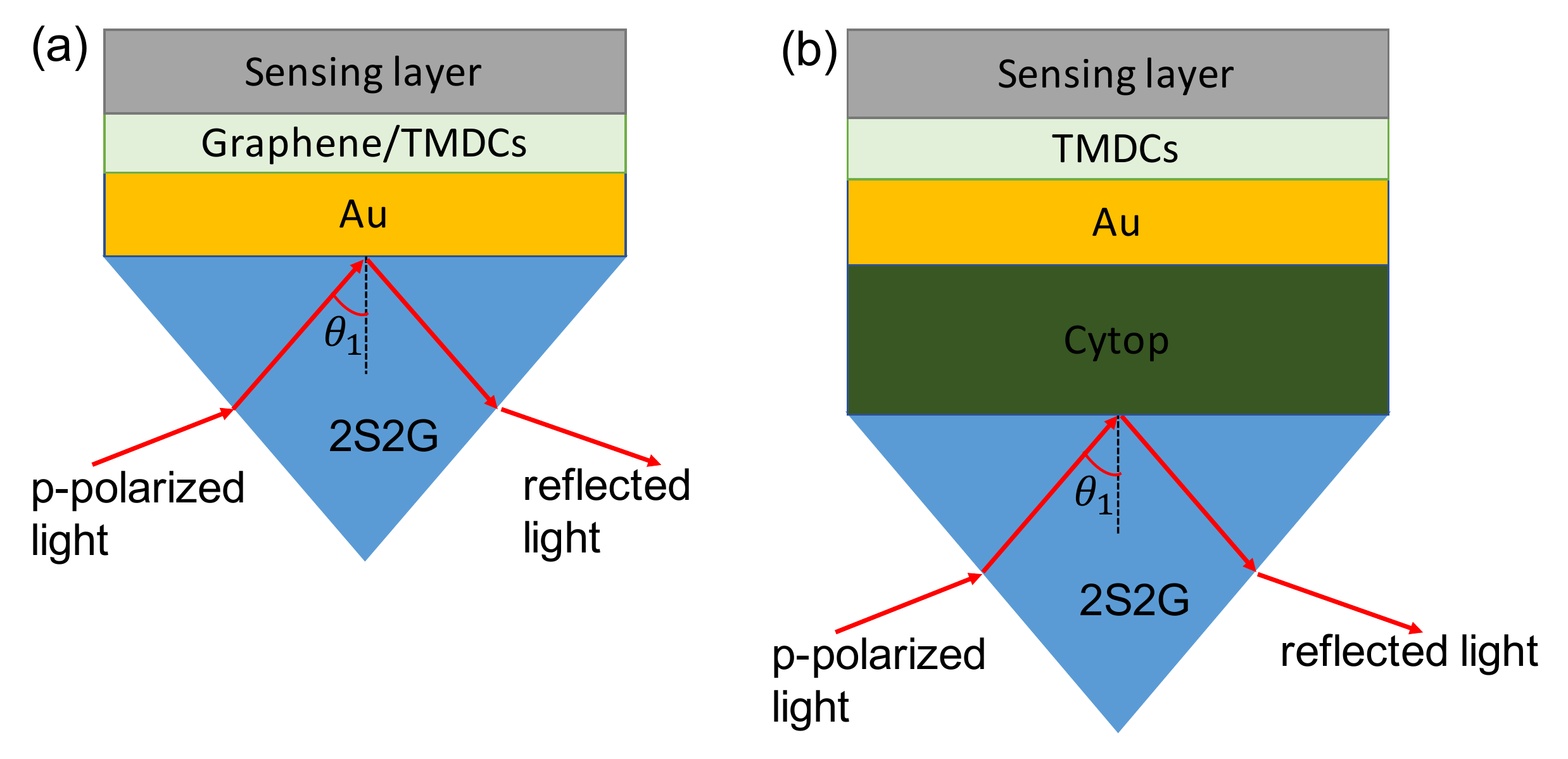}     
      \caption{Schematic diagram of of 2D material-based (a) conventional SPR sensor and (b) LRSPR sensor.}
\label{fig1}
\end{figure}
\begin{figure}[thpb]
      \centering
      \includegraphics[scale=0.26]{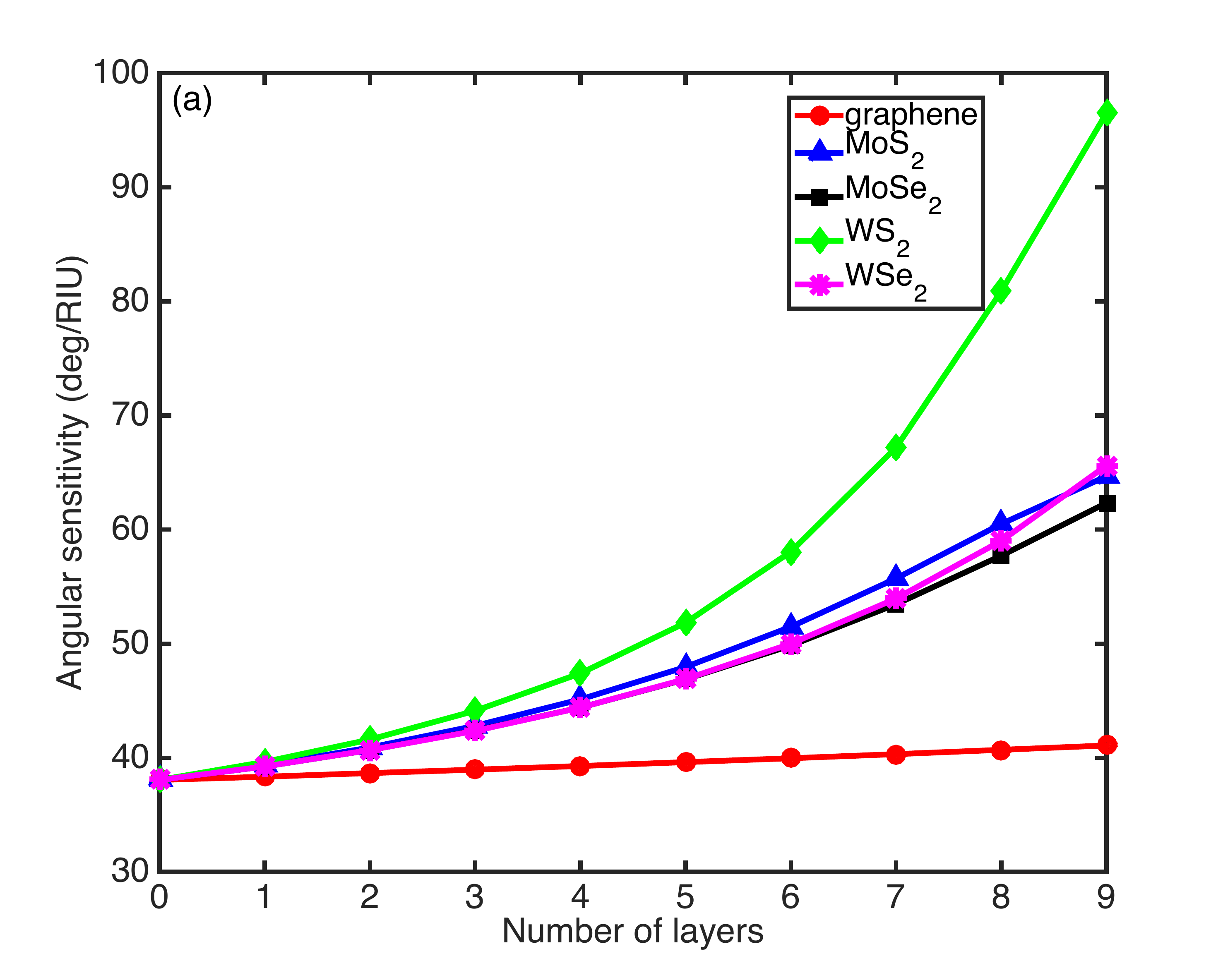}
      \includegraphics[scale=0.26]{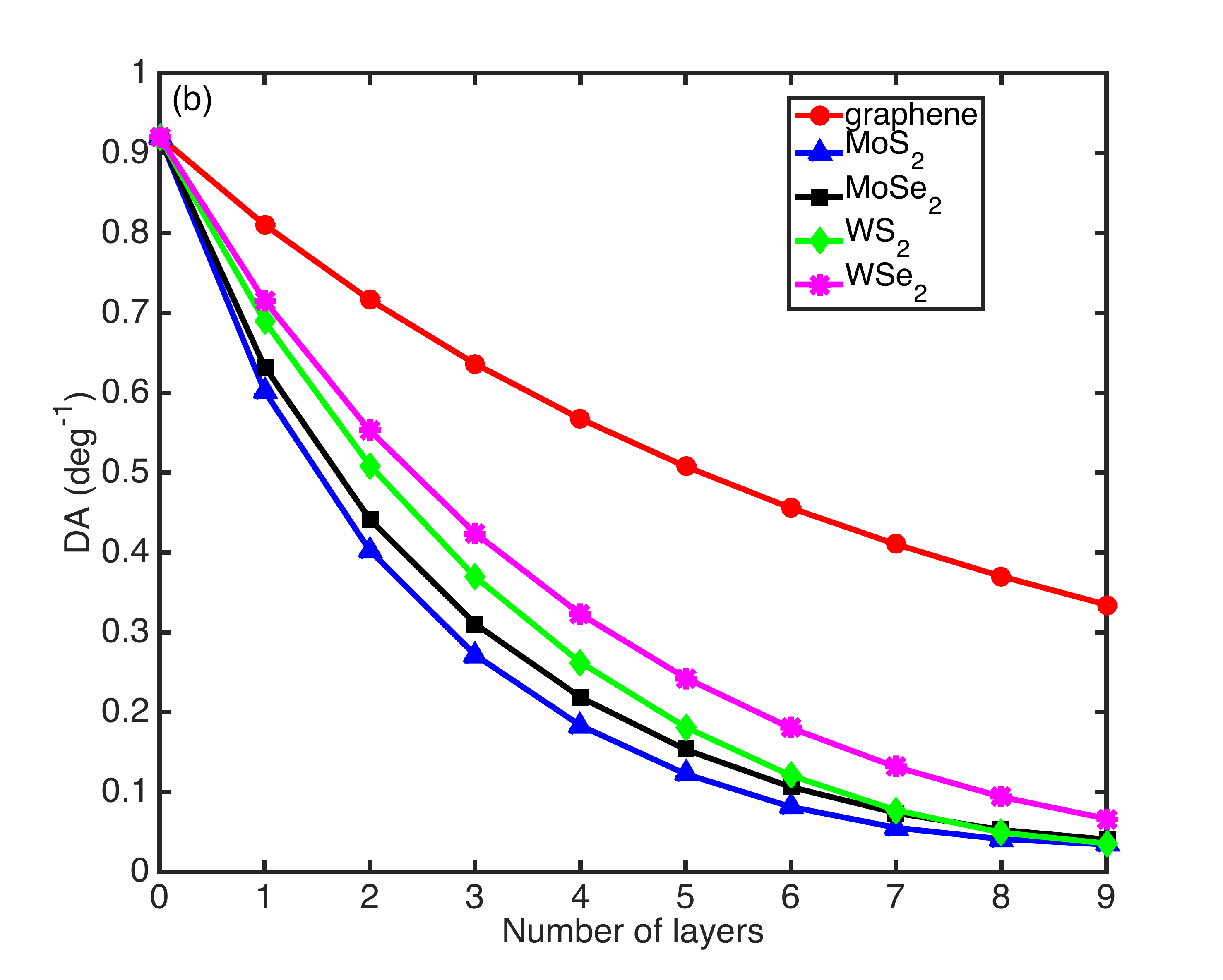}         
      \caption{(a) Angular sensitivity and (b) DA of graphene- and TMDCs-based cSPR sensor. The thickness of Au thin film is 50 nm.}
\label{fig2}
\end{figure}
\subsection{Angle modulated SPR sensor}
Firstly, we compare the angular sensitivity of graphene- and TMDCs-based cSPR sensor with angle modulation. The prism-coupled cSPR sensor is shown in Fig. \ref{fig1}(a), in which graphene or TMDCs coated Au thin film is attached to the 2S2G prism. A p-polarized light beam with fixed wavelength $\lambda=633$ nm is employed to excite the SPWs, and the reflectance-angle curve is obtained by scanning the incident angle. The angular sensitivity $S_{angle}$ for cSPR sensors with graphene and four TMDC materials (${\text{MoS}}_2$, ${\text{MoSe}}_2$, ${\text{WS}}_2$, ${\text{WSe}}_2$) is shown in Fig. \ref{fig2}(a). For graphene- and TMDCs-based cSPR sensor, the angular sensitivity increases with the number of 2D material layers. This enhancement can be attributed to the charge transfer from 2D material to the surface of Au thin film \cite{giovannetti2008doping,zeng2014nanomaterials}, which in turn enhances the electric field distribution at the sensing surface. However, graphene cSPR sensor shows lower angular sensitivity and sensitivity growth rate than that of TMDCs-based cSPR sensor. ${\text{WS}}_2$ shows the highest angular sensitivity, and a sensitivity of $\sim100$ deg/RIU was obtained with 9 layers ${\text{WS}}_2$, while $\sim41$ deg/RIU for 9 layers graphene-based cSPR sensor. This indicates that TMDCs can be a good candidate for RI sensing applications. However, these cSPR sensors exhibits a low DA ($<1\ {\text{deg}}^{-1}$) and decreases with the number of layers.  
\begin{figure}[thpb]
      \centering
      \includegraphics[scale=0.27]{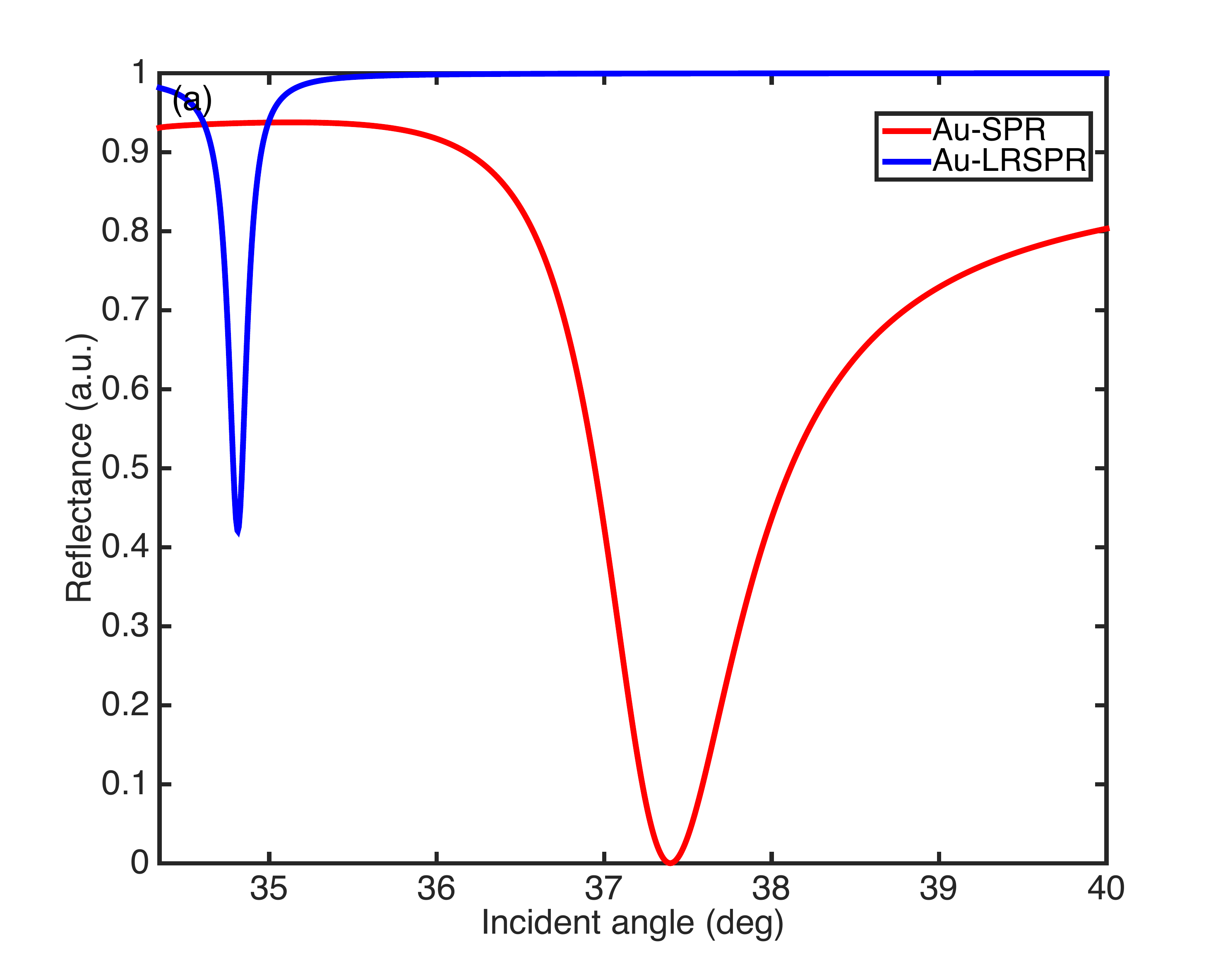}
      \includegraphics[scale=0.27]{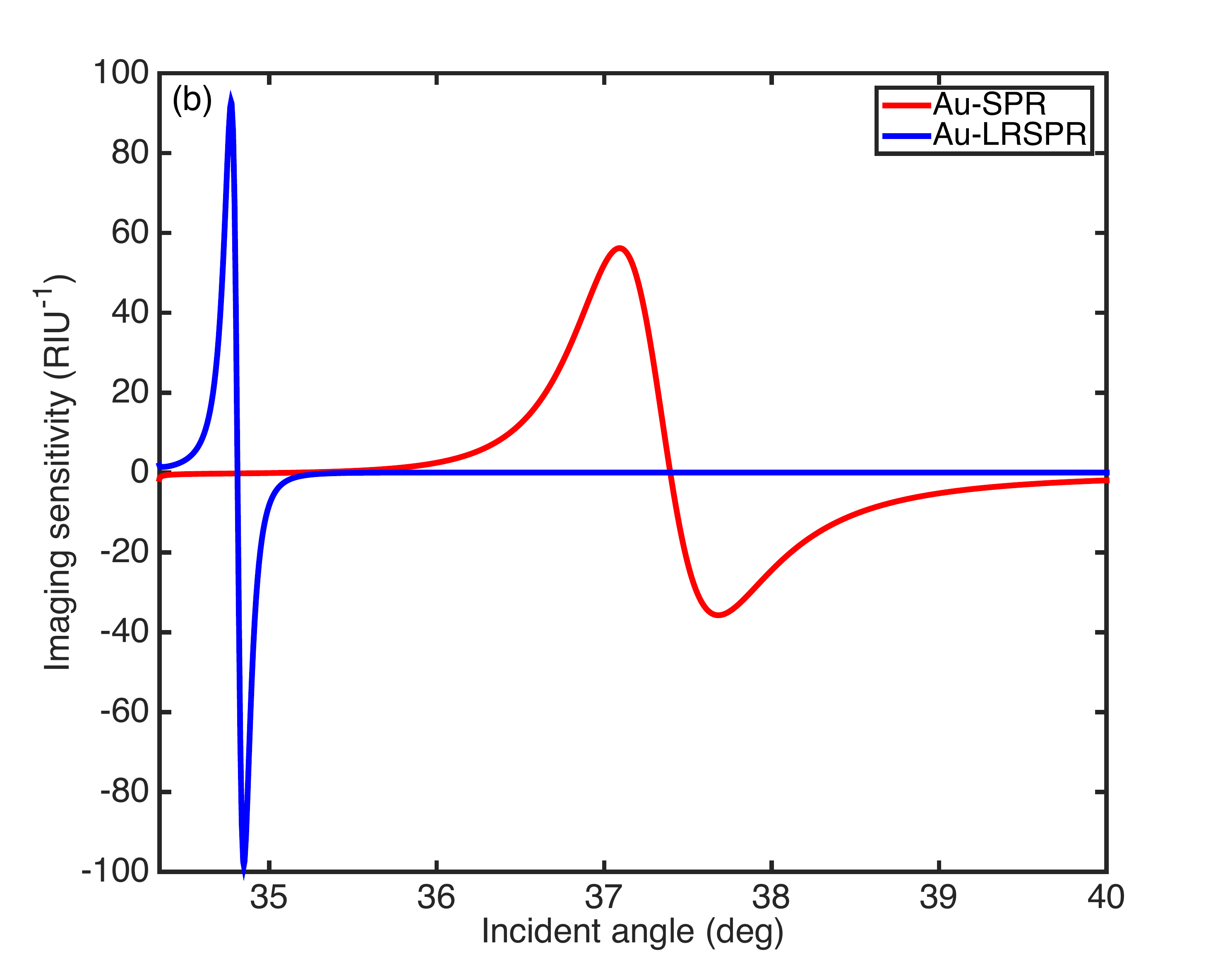}  
      \includegraphics[scale=0.36]{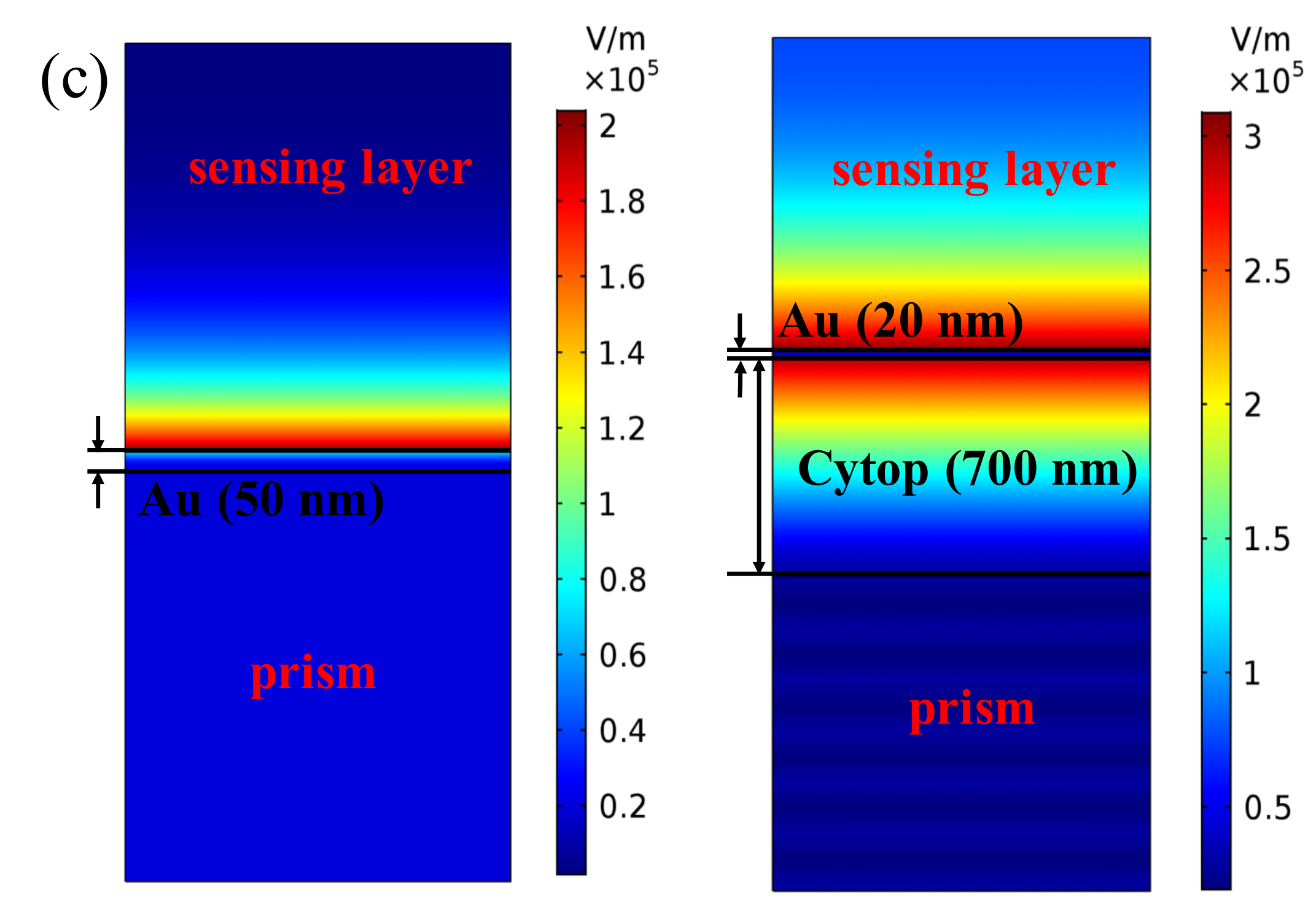}
      \includegraphics[scale=0.36]{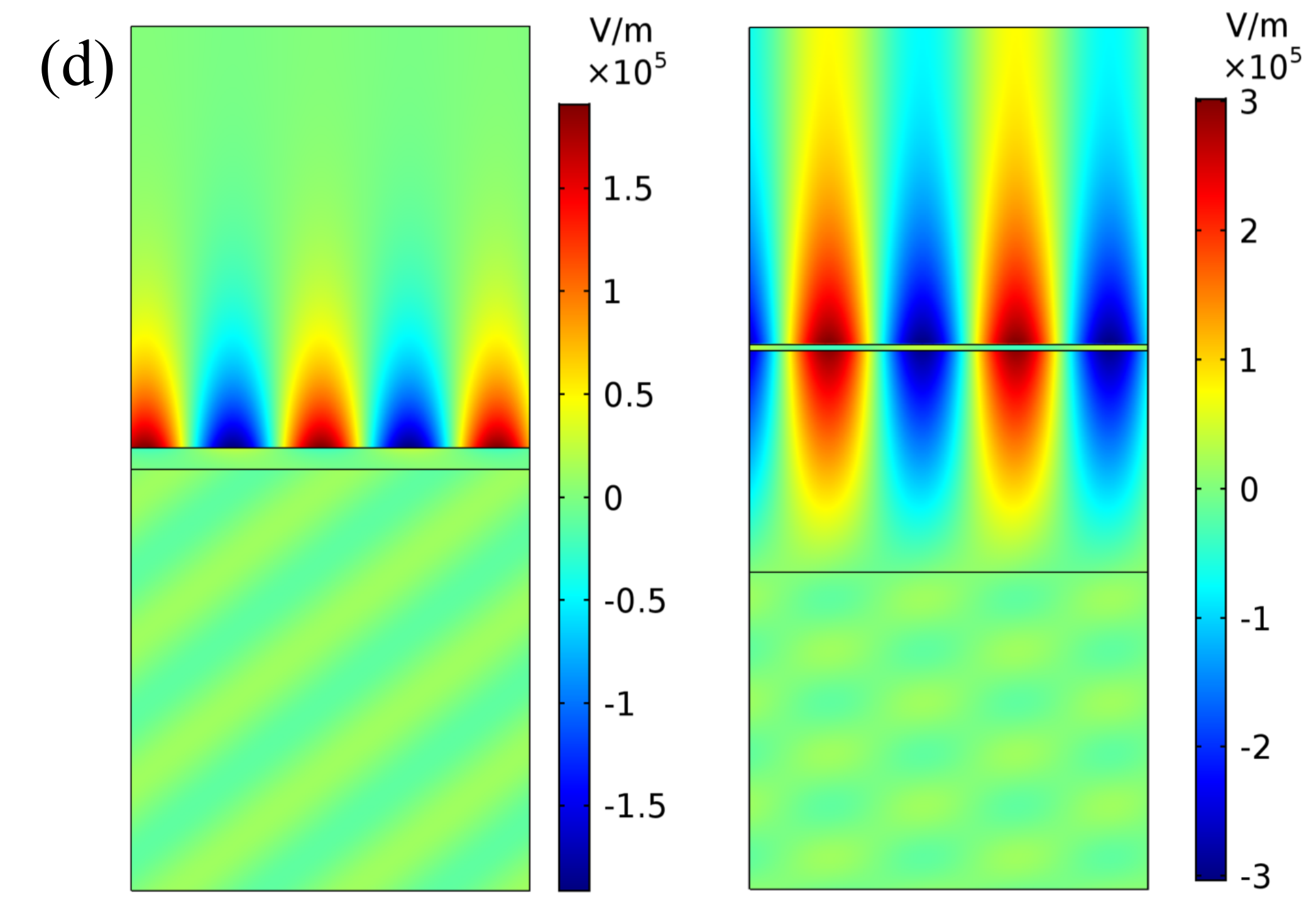}         
      \caption{(a) Reflectance and (b) imaging sensitivity as a function of the incident angle for Au-based cSPR and LRSPR sensor without any coating layer. Distribution of (c) electric field norm and (d) $y$ components of the electric field for cSPR (left) and LRSPR (right) sensor at the resonance angle. The Au film thickness for cSPR sensor and LRSPR sensor are 50 nm and 20 nm, respectively. The thickness of cytop layer in LRSPR sensor is 0.7 $\mu$m.}
\label{fig3}
\end{figure}

To overcome the low DA of TMDCs-based cSPR sensor, we proposed a LRSPR sensor (see Fig. \ref{fig1}(b)), in which a cytop layer was embedded between the 2S2G prism and the TMDCs coated Au thin film. As shown in Fig. \ref{fig3}(a), the LRSPR sensor with 20 nm thick Au film shows narrower reflectance-angle curve (i.e., smaller FWHM) than that of cSPR with 50 nm Au film (typical metal thickness in cSPR sensor), which in turn improves the DA of sensor. In cSPR sensor, SPWs are excited at the interface of metal-sensing layer, while SPWs are exist at both interfaces of Au-sensing layer and Au-cytop layer (see Figs. \ref{fig3}(c)-(d)). The coupling of these two SPWs gives rise to LRSPR, which shows larger field penetration into sensing layer (545 nm vs. 189nm for cSPR). Typical thickness of the Au film for LRSPR is 15 - 30 nm to provide a strong coupling between the two SPWs. In addition, LRSPR has more flexibility to tune cytop thickness to optimize the sensing performance. The imaging sensitivities of cSPR and LRSPR sensor are shown in Fig. \ref{fig3}(b), which exhibit positive and negative peaks. Here, for convenience we only consider the positive peak imaging sensitivity which is referred to as imaging sensitivity in the following. It is found that the LRSPR sensor exhibits higher imaging sensitivity than that of cSPR sensor. Therefore, in the following, we will focus on the imaging sensitivity and DA of TMDCs-based LRSPR imaging biosensor. 

\section{Results and discussion}
\begin{figure}[thpb]
      \centering
      \includegraphics[scale=0.27]{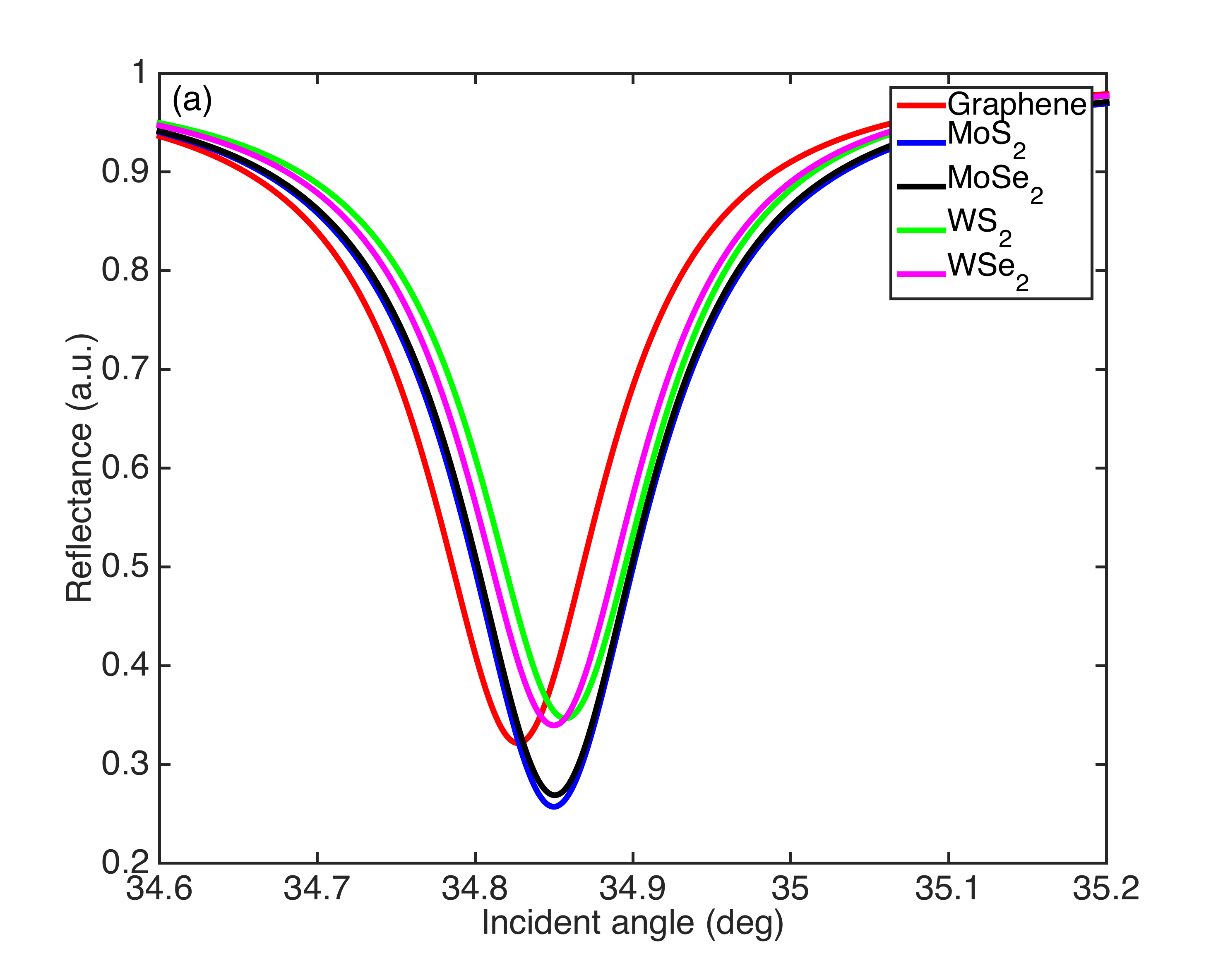}
      \includegraphics[scale=0.27]{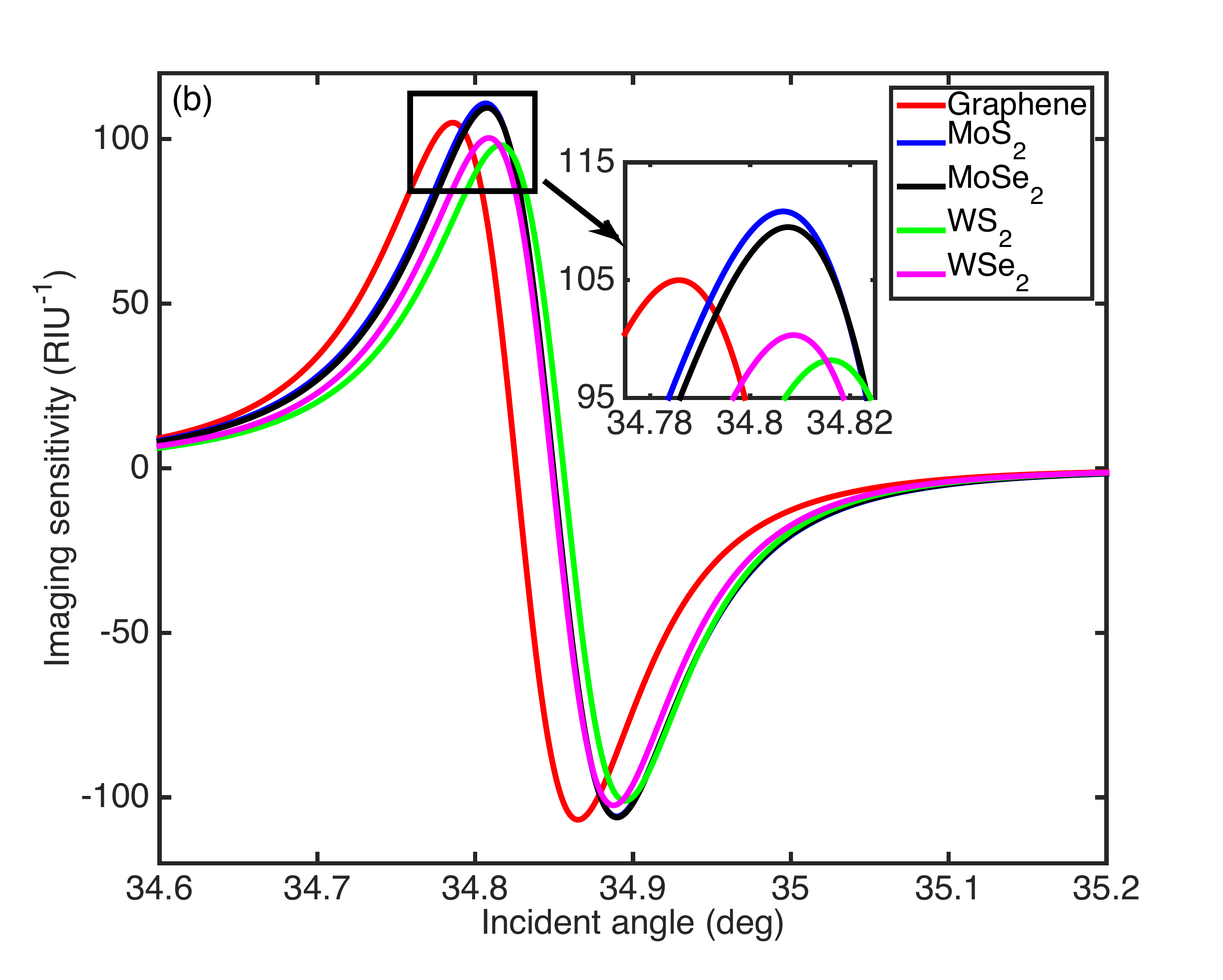}         
      \caption{(a) Reflectance and (b) imaging sensitivity as a function of the incident angle for LRSPR sensor based on graphene-on-Au and TMDCs-on-Au structure. The Au film thickness for cSPR sensor and LRSPR sensor are 50 nm and 20 nm, respectively. The thickness of cytop layer in LRSPR sensor is 0.7 $\mu$m.}
\label{fig4}
\end{figure}
The reflectance and imaging sensitivity of graphene- and TMDCs-based LRSPR imaging sensor are shown in Figs. \ref{fig4}(a) and (b), respectively. It is found that the imaging sensitivity has improved with the combination of 2D materials (graphene and TMDCs) and Au film. However, these 2D materials LRSPR sensors show degraded DA as compared to that of Au-based LRSPR sensor (see Table. \ref{table1}), which is a result of the increased damping surface plasmon oscillations with the introduction of absorbing 2D materials. The LRSPR sensor with ${\text{MoS}}_2$-on-Au structure has the highest imaging sensitivity and the lowest DA. Despite this decreased DA, it exhibits more than 7-fold enhancement in comparison with that of 2D material-based cSPR sensor (see Fig. \ref{fig2}(b)). 
 
\begin{table}[thpb]
\caption{\label{table1}%
Sensor performance of graphene- and TMDCs-based LRSPR sensor
}
\begin{ruledtabular}
\begin{tabular}{ccc}
\textrm{Material}& \textrm{Sensitivity} & \textrm{DA}\\
\colrule
 Au & 92.82 & 7.82\\
 Graphene-Au & 105 & 7.45 \\
 ${\text{MoS}}_2$-Au & 110.89  & 7.10 \\
 ${\text{MoSe}}_2$-Au & 109.50  & 7.17\\
 ${\text{WS}}_2$-Au & 98.19 & 7.50 \\
  ${\text{WSe}}_2$-Au & 100.33 & 7.49 \\
\end{tabular}
\end{ruledtabular}
\end{table}

\begin{figure}[thpb]
      \centering
      \includegraphics[scale=0.265]{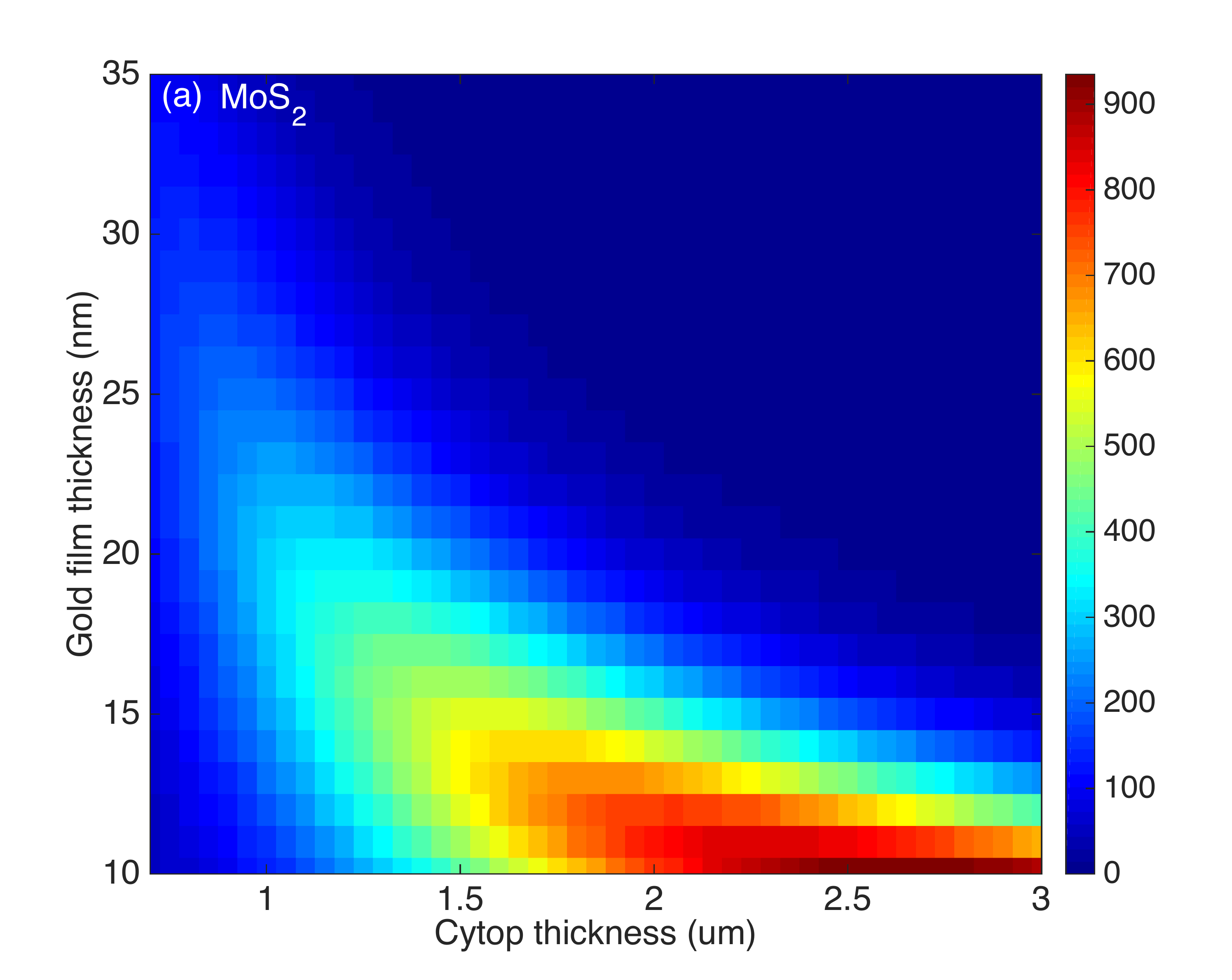}
      \includegraphics[scale=0.265]{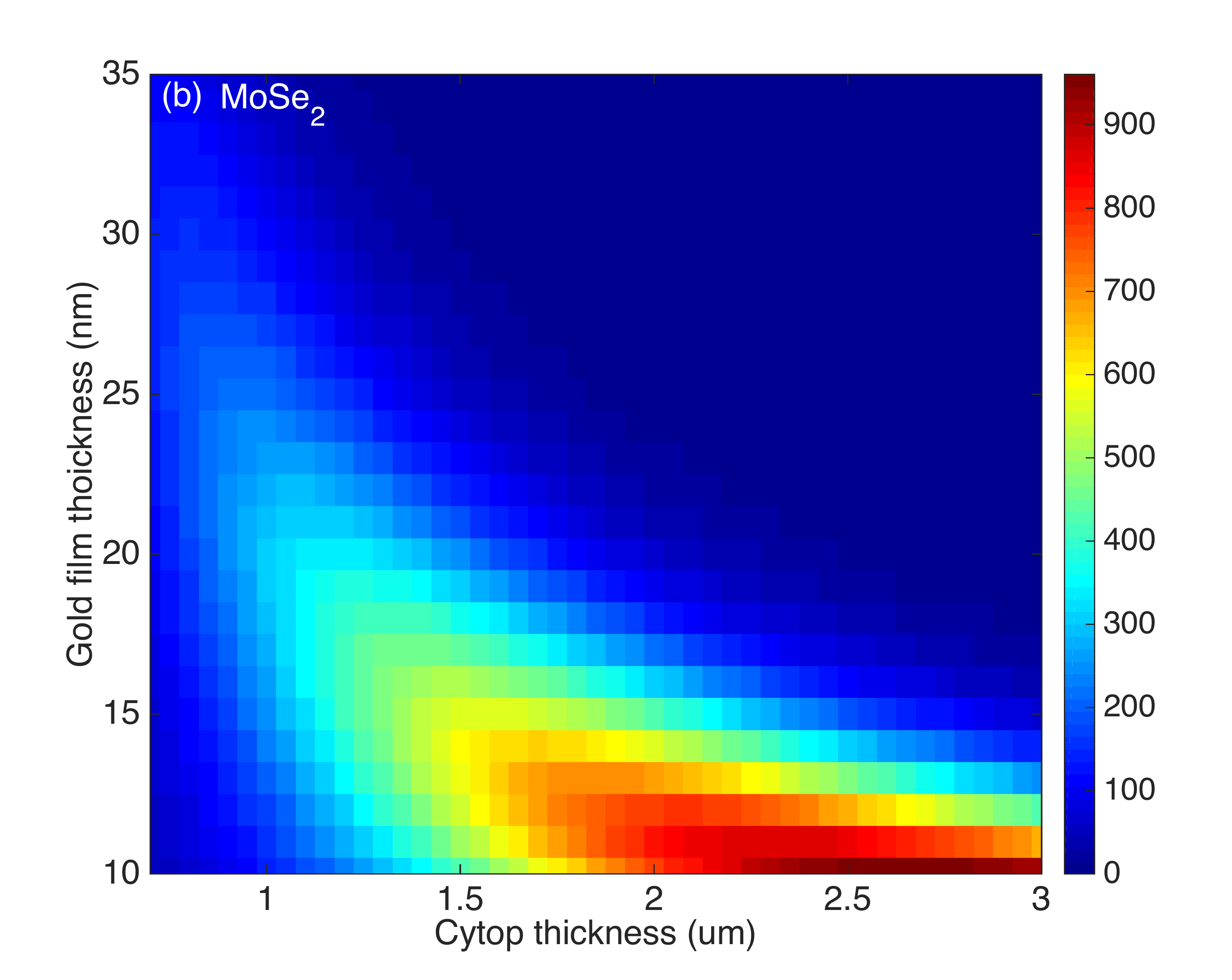}  
      \includegraphics[scale=0.265]{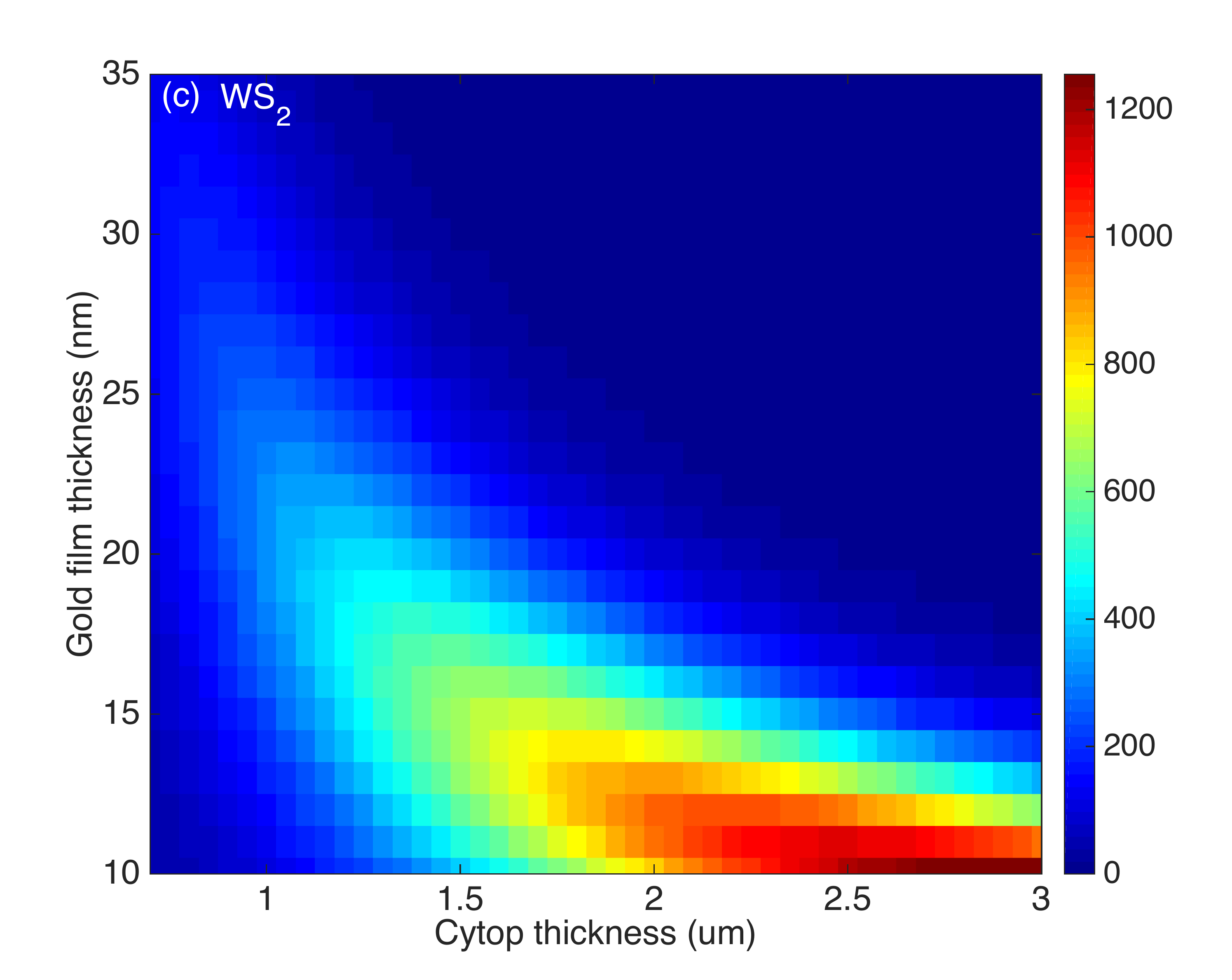}
      \includegraphics[scale=0.265]{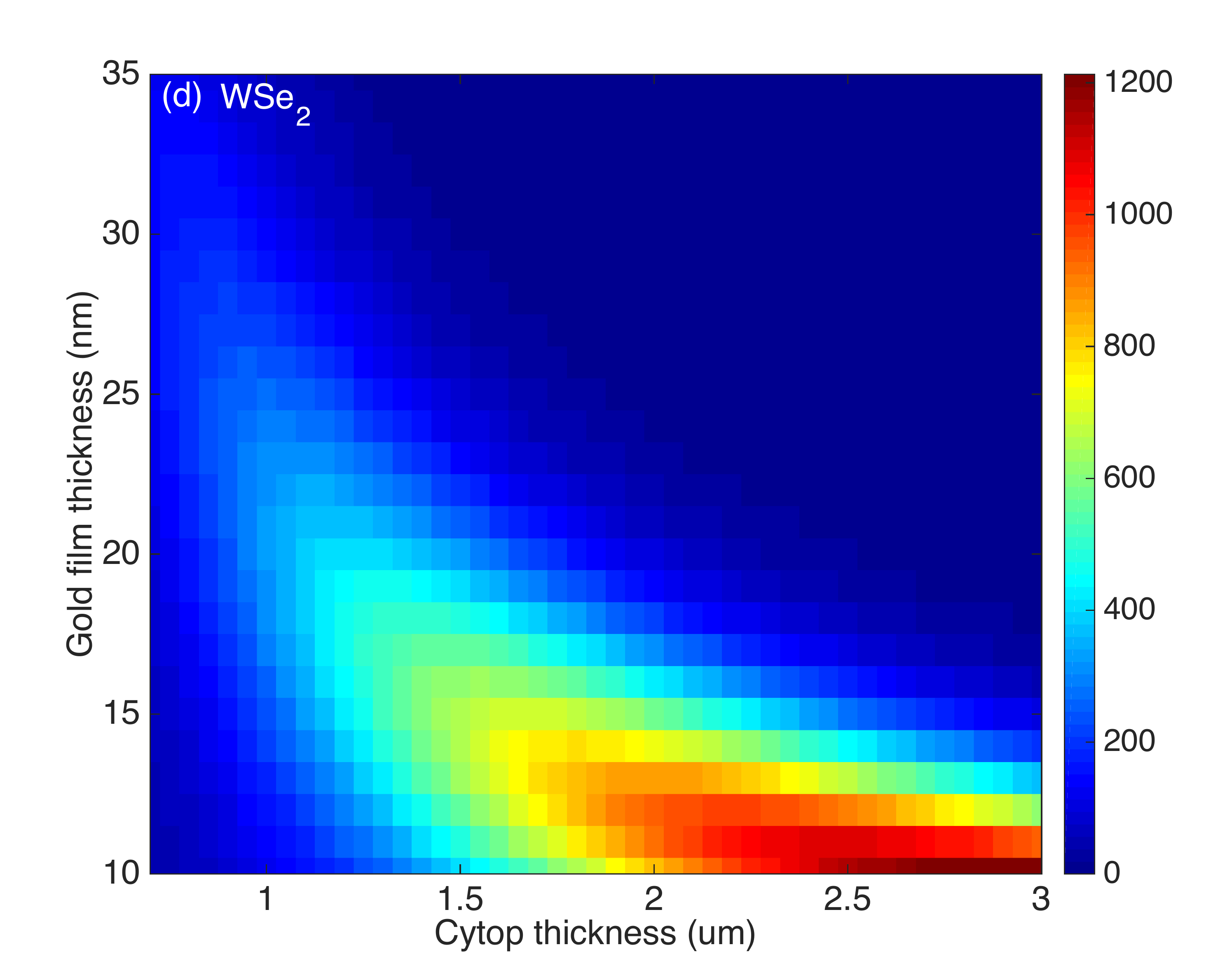}         
      \caption{Imaging sensitivity as a function of the thickness of Au film and cytop layer for LRSPR with (a) ${\text{MoS}}_2$, (b) ${\text{MoSe}}_2$, (c) ${\text{WS}}_2$, (d) ${\text{WSe}}_2$.}
\label{fig5}
\end{figure}

\begin{figure}[thpb]
      \centering
      \includegraphics[scale=0.265]{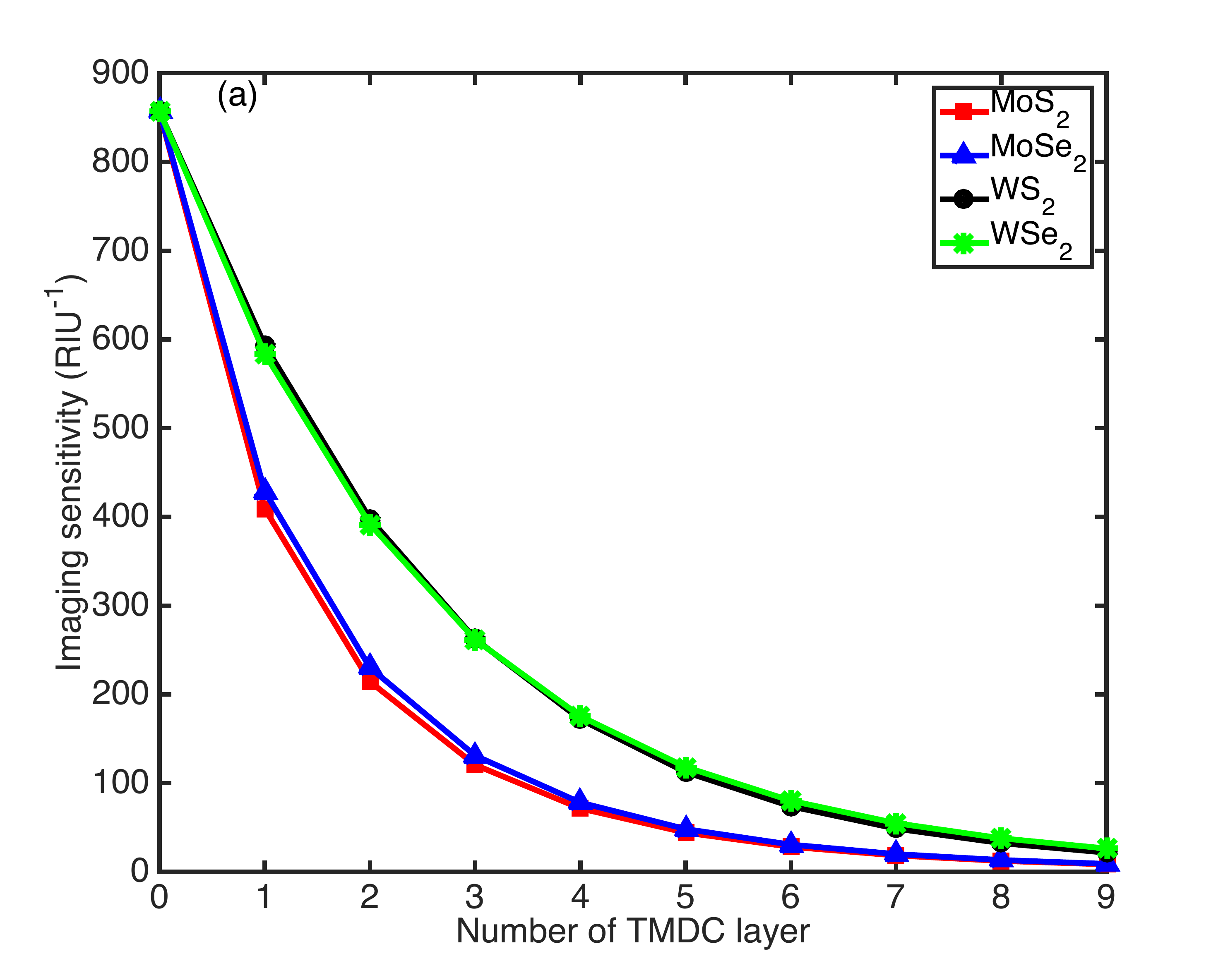}
      \includegraphics[scale=0.265]{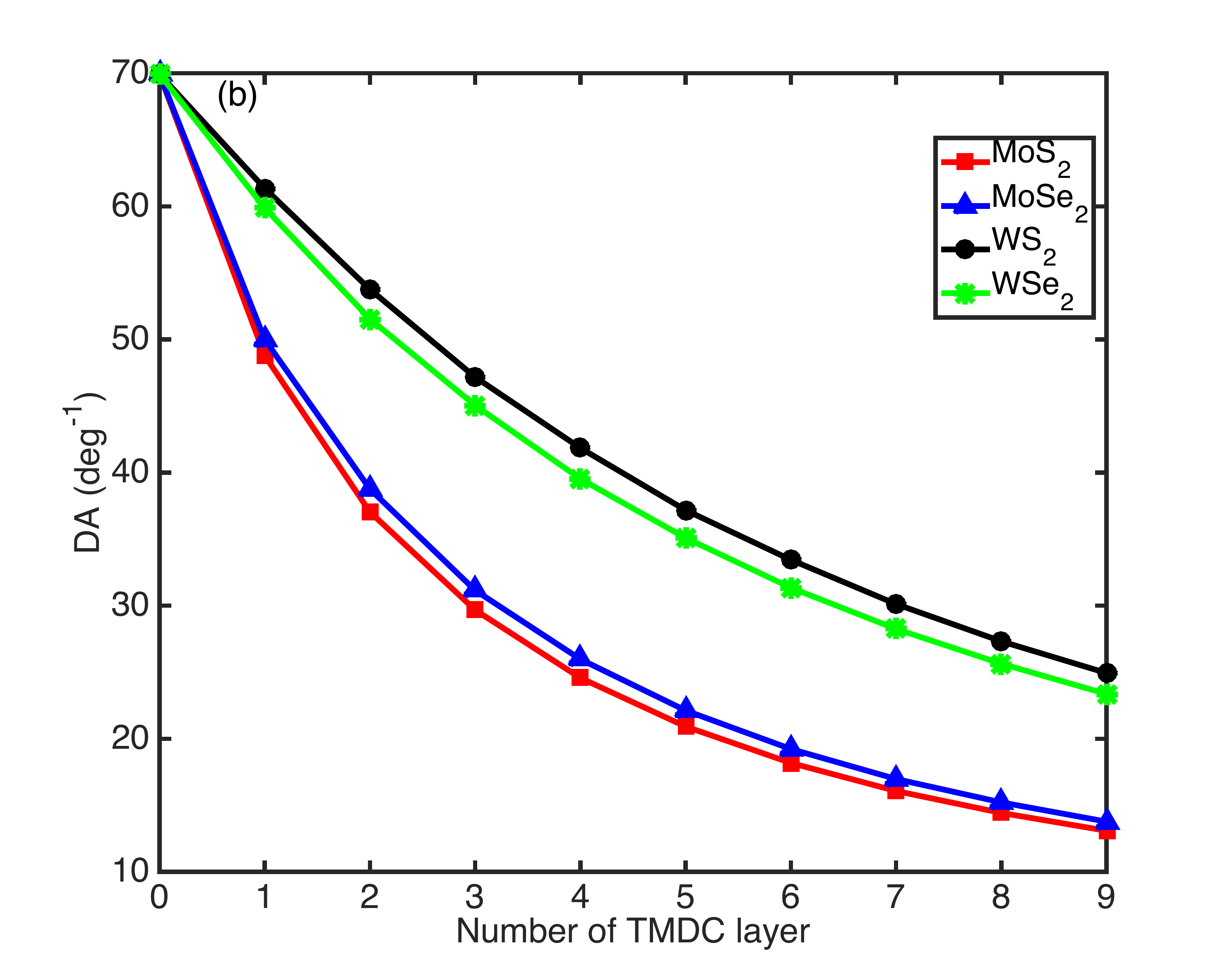}  
      \includegraphics[scale=0.265]{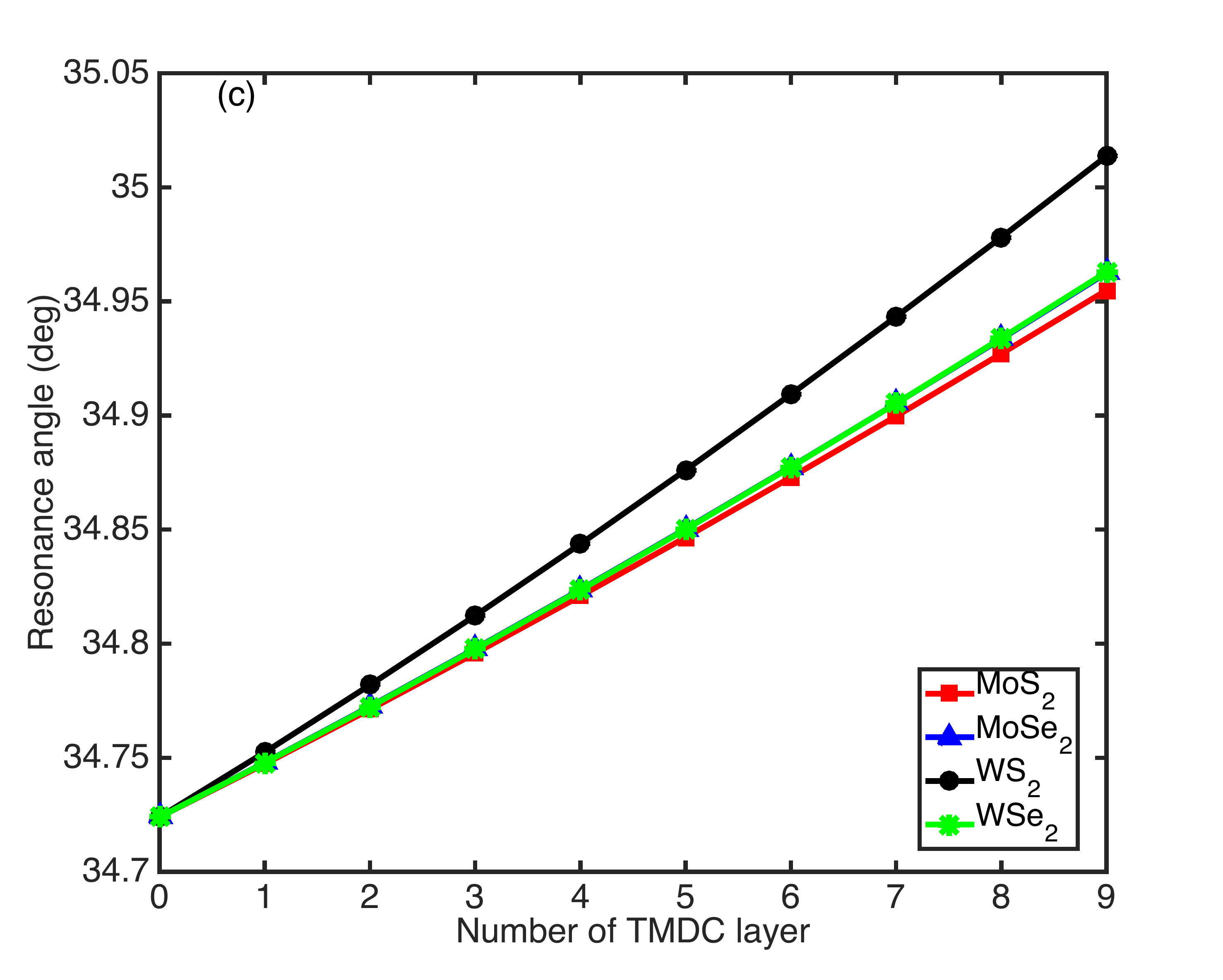}
      \includegraphics[scale=0.265]{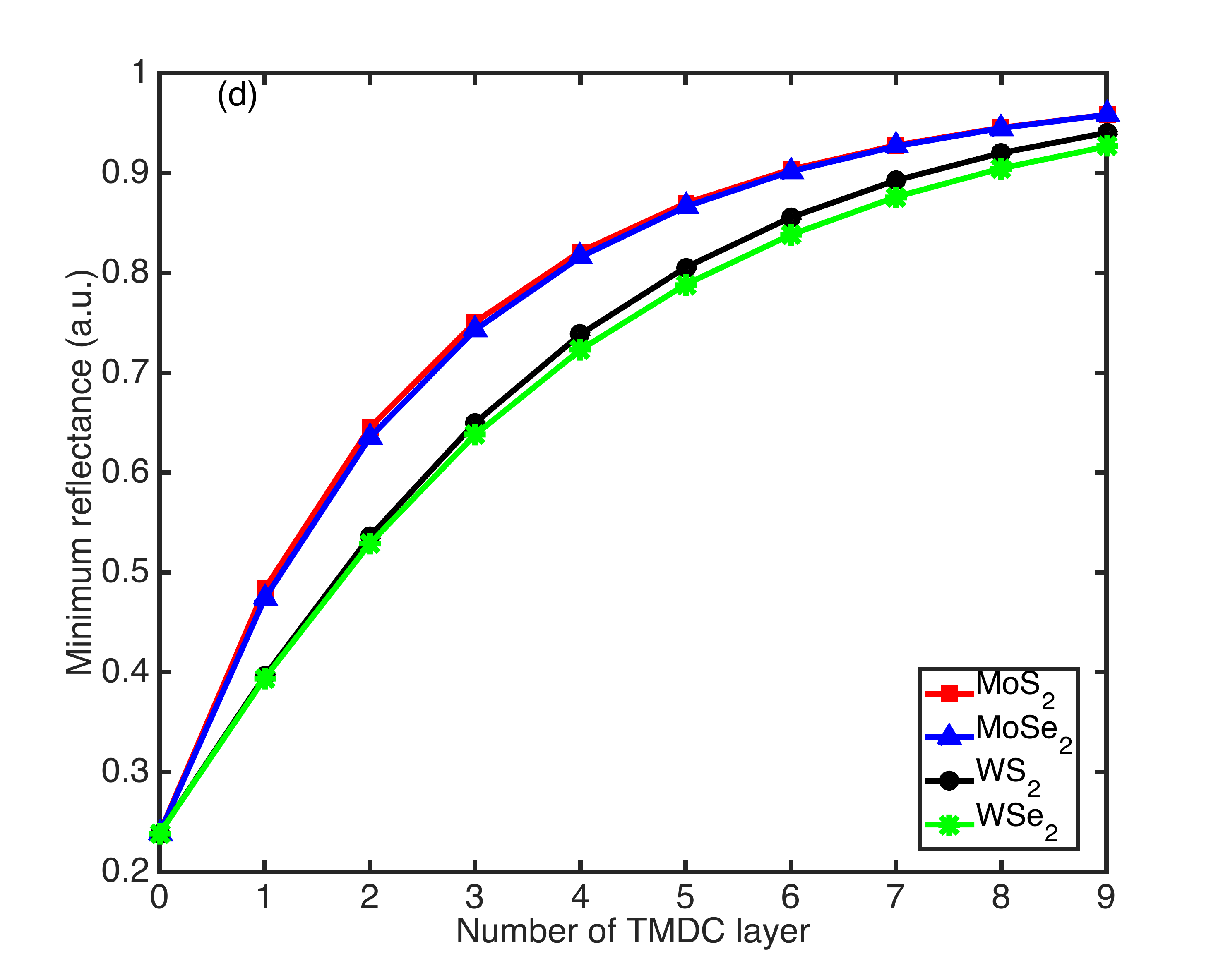}         
      \caption{(a) Imaging sensitivity, (b) DA, (c) resonance angle, and (d) minimum reflectance for LRSPR sensor with multiple TMDC layers. The thickness of Au thin film and cytop layer are 15 nm and 2.0 $\mu$m, respectively.}
\label{fig6}
\end{figure}
\begin{figure}[thpb]
      \centering
      \includegraphics[scale=0.265]{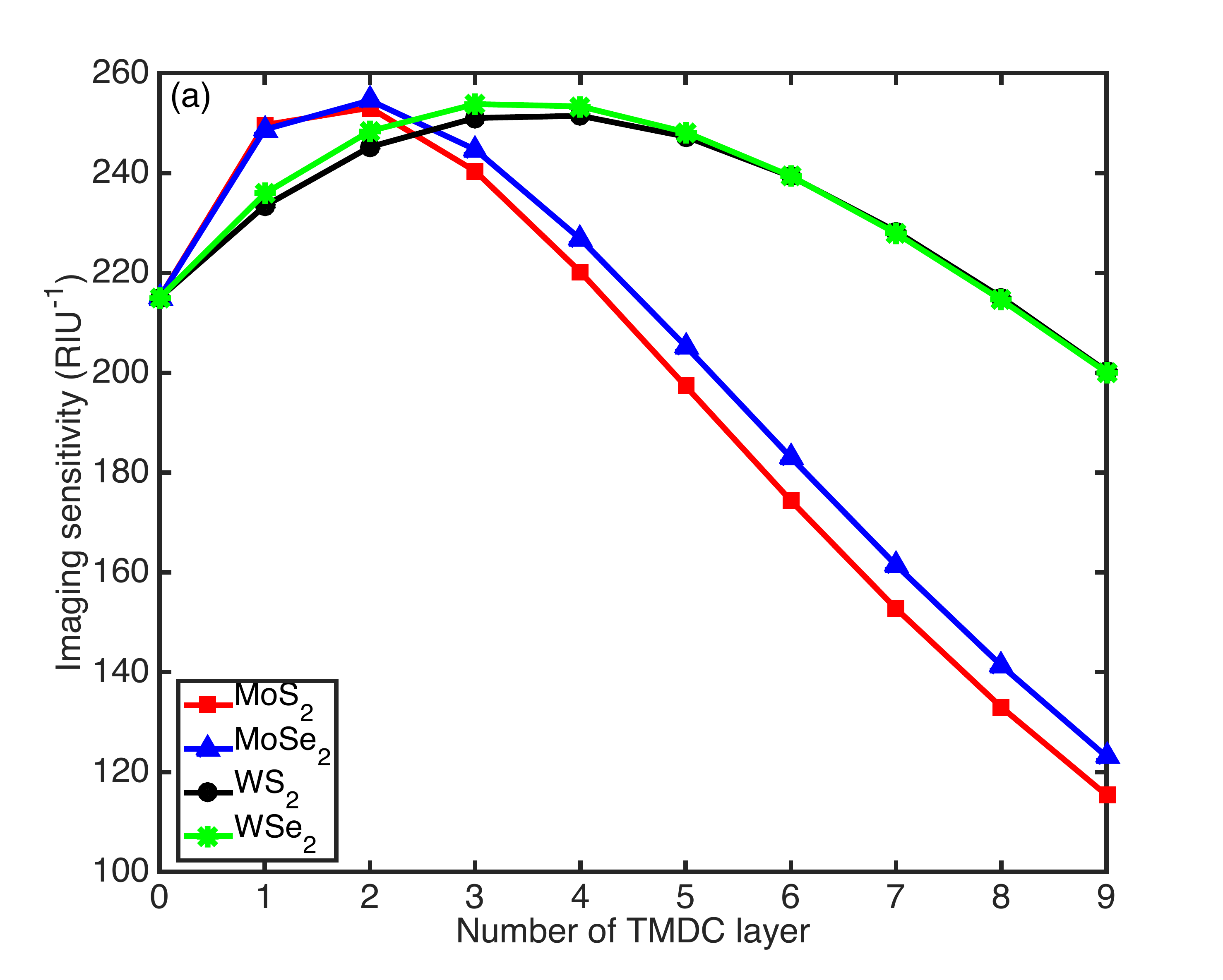}
      \includegraphics[scale=0.265]{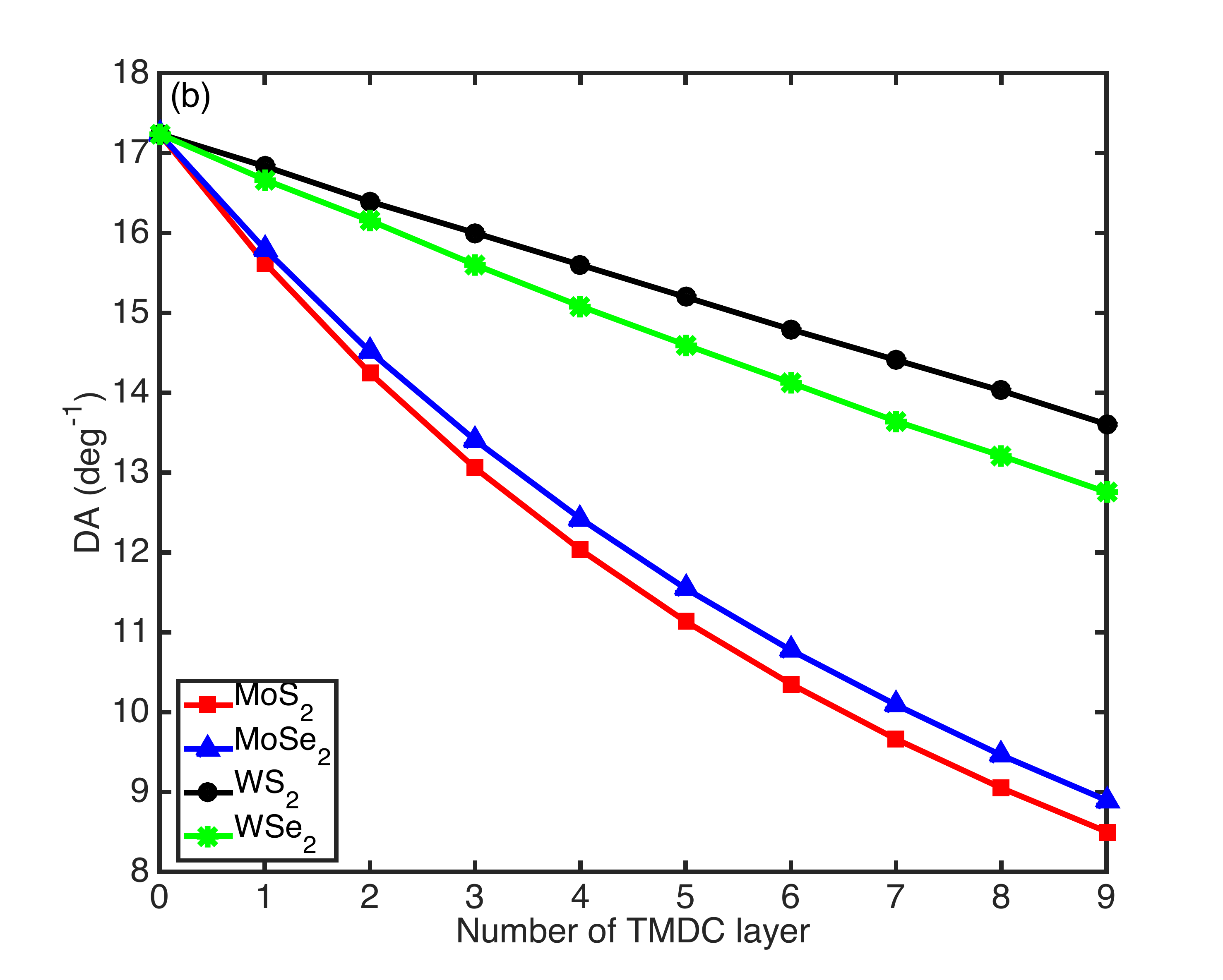}  
      \includegraphics[scale=0.265]{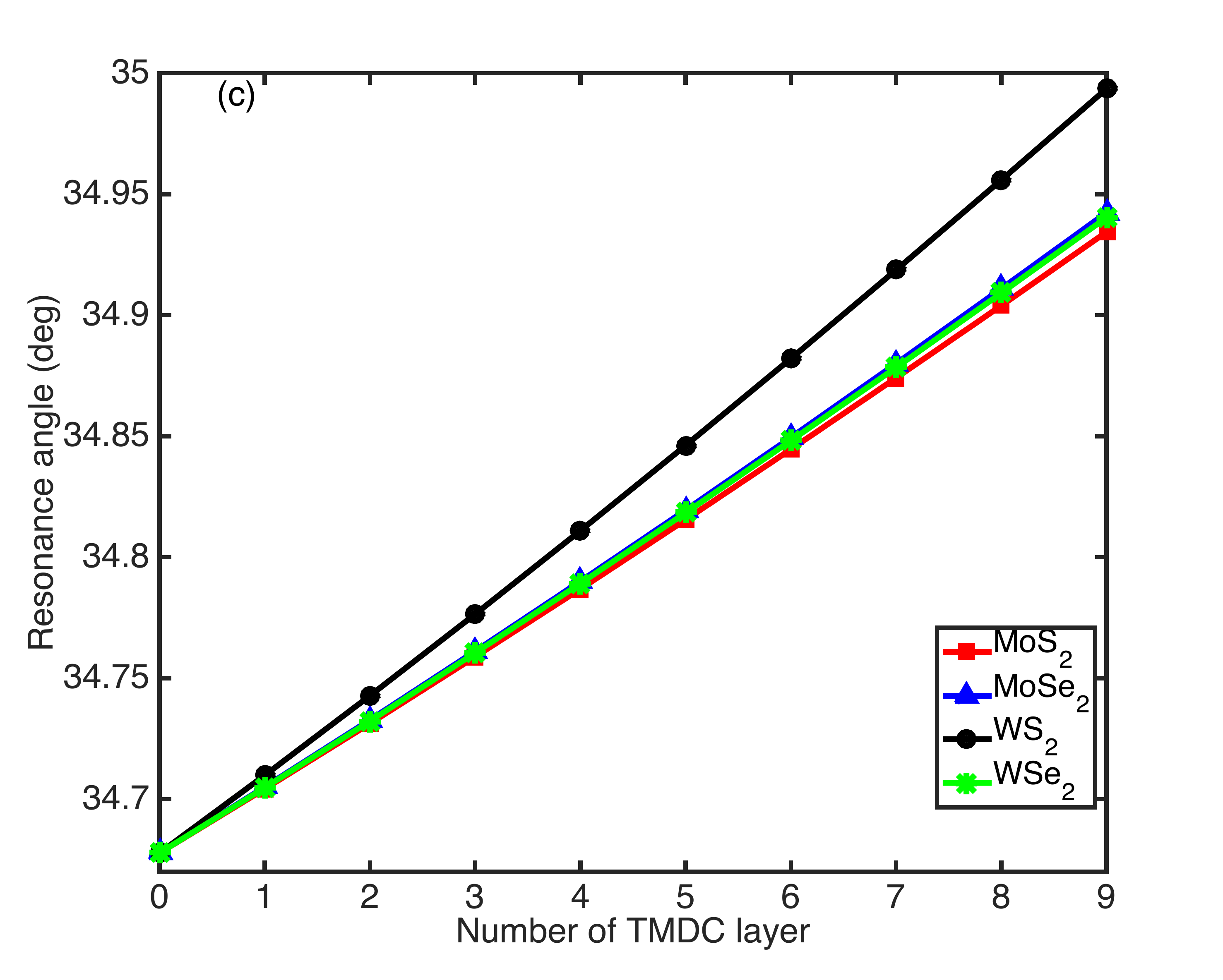}
      \includegraphics[scale=0.265]{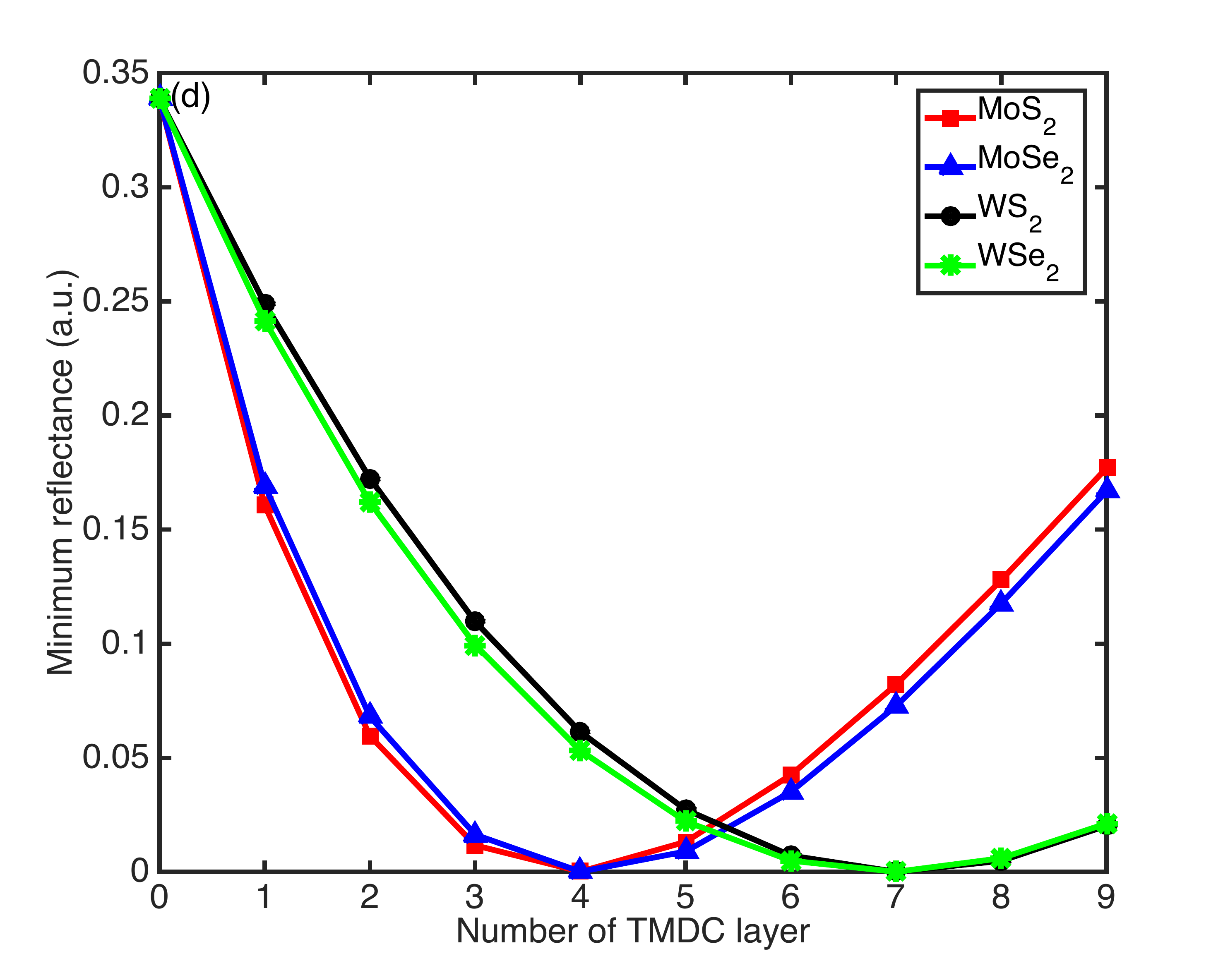}         
      \caption{(a) Imaging sensitivity, (b) DA, (c) resonance angle, and (d) minimum reflectance for LRSPR sensor with multiple TMDC layers. The thickness of Au thin film and cytop layer are 15 nm and 1.0 $\mu$m, respectively.}
\label{fig7}
\end{figure}

The study above only focuses on one particular Au and Cytop thickness. For LRSPR configuration, it is possible to tune the thickness of Au film ($d_{\text{Au}}$) and cytop layer ($d_{\text{cytop}}$) to optimize the sensor performance. The imaging sensitivity of the LRSPR sensor as a function of $d_{\text{Au}}$ and $d_{\text{cytop}}$ is shown in Fig. \ref{fig5}. It is found that the Au film should be thin to get a higher imaging sensitivity with a thick cytop layer. Au film with thickness $>25$ nm is not a good choice in obtaining a relative high imaging sensitivity. In general, the Tungsten-based TMDC (${\text{WS}}_2$ and ${\text{WSe}}_2$) LRSPR sensor exhibits higher imaging sensitivity than that of Molybdenum-based TMDC (${\text{MoS}}_2$ and ${\text{MoSe}}_2$) sensor. For example, ${\text{WS}}_2$- and ${\text{WSe}}_2$-based LRSPR sensors show a maximum sensitivity of more than 1200 $\text{RIU}^{-1}$, while the highest sensitivity for ${\text{MoS}}_2$ and ${\text{MoSe}}_2$ LRSPR sensors are $\sim$935 $\text{RIU}^{-1}$ and $\sim$960 $\text{RIU}^{-1}$, respectively. In addition to the sensitivity enhancement with the employment of TMDCs in LRSPR sensor (see Table \ref{table1}), we also found that the TMDC layers degrades the LRSPR sensor sensitivity as compared to the TMDCs-devoid setup. The imaging sensitivity enhanced and degraded effect depend on the thickness of Au film and cytop layer. For example, monolayer TMDC-based LRSPR sensors exhibit lower sensitivity than that of Au-based LRSPR sensor with $d_{\text{Au}}=15$ nm and $d_{\text{cytop}}=2.0$ (degraded sensitivity, case 1), as shown in Fig. \ref{fig6}(a). However, it can be seen from Fig. \ref{fig7}(a) that the imaging sensitivity can be improved with the employment of monolayer TMDC for $d_{\text{Au}}=15$ nm and $d_{\text{cytop}}=1.0$ $\mu\text{m}$ (enhanced sensitivity, case 2). This imaging sensitivity enhancement for the proposed LRSPR imaging biosensor is different from the previous graphene-based LRSPR or cSPR imaging sensor, in which the imaging sensitivity are usually decreases with number of graphene layers \cite{choi2011graphene,maharana2014performance,maharana2013ultrasensitive,maharana2014low,wu2016long} (similar to the case 1). In the following, we will investigate the LRSPR sensor performance in the two cases. 
 
The effect of multiple TMDC layers on the imaging sensitivity for case 1 is shown in Fig. \ref{fig6}(a). In this case, the imaging sensitivity decreases with the number of TMDC layers, which exhibits lower sensitivity than that of Au-based LRSPR sensor. The degradation of sensitivity can be attributed to the non-zero imaginary part of RI which in turn increases the plasmon damping with multiple TMDC layers. The DA for LRSPR sensor (Fig. \ref{fig6}(b)) deceases with the number of TMDC layers due to the broader reflectance-angle curve, which is a result of the presence of absorbing TMDC layers. Here, $\text{MoS}_2$-based biosensor has the minimum DA, while $\text{WS}_2$ biosensor exhibits the maximum DA. This is because that $\text{WS}_2$ has a minimum imaginary part of the complex RI at wavelength $\lambda=633$ nm, while $\text{MoS}_2$ has the maximum imaginary part which causes a lowest DA. In addition, with the increasing number of TMDC layers, the resonance angle shifts towards higher incident angle, and the reflectance-angle curve of LRSPR sensor becomes shallower (i.e., higher minimum reflectance), as shown in Figs. \ref{fig6}(c) and (d), respectively. For case 2, the imaging sensitivity first increases with the number of TMDC layers and then decreases, as shown in Fig. \ref{fig7}(a). An improved imaging sensitivity can be obtained for few layers of TMDC as compared to bare Au-based LRSPR sensor. It is found that the enhancement of imaging sensitivity can persist for 4 layers of $\text{MoS}_2$ and $\text{MoSe}_2$ sheet, while 7 layers of $\text{WS}_2$ and $\text{WSe}_2$ sheet. Similar to case 1, the DA and resonance angle decreases and increases with the number of TMDC layers, as shown in Figs. \ref{fig7}(b) and (c), respectively. Although the DA for LRSPR sensor in case 2 is lower than that of case 1, it is still much higher than that of cSPR sensor. For the minimum reflectance of reflectance-angle curve in case 2, it first decreases and the increases with the number TMDC layers, and exhibits a minimum value at 4 layers $\text{MoS}_2$ and $\text{MoSe}_2$, while 7 layers for $\text{WS}_2$ and $\text{WSe}_2$ sheet (Fig. \ref{fig7}(d)).  
   
\begin{figure}[thpb]
      \centering
      \includegraphics[scale=0.27]{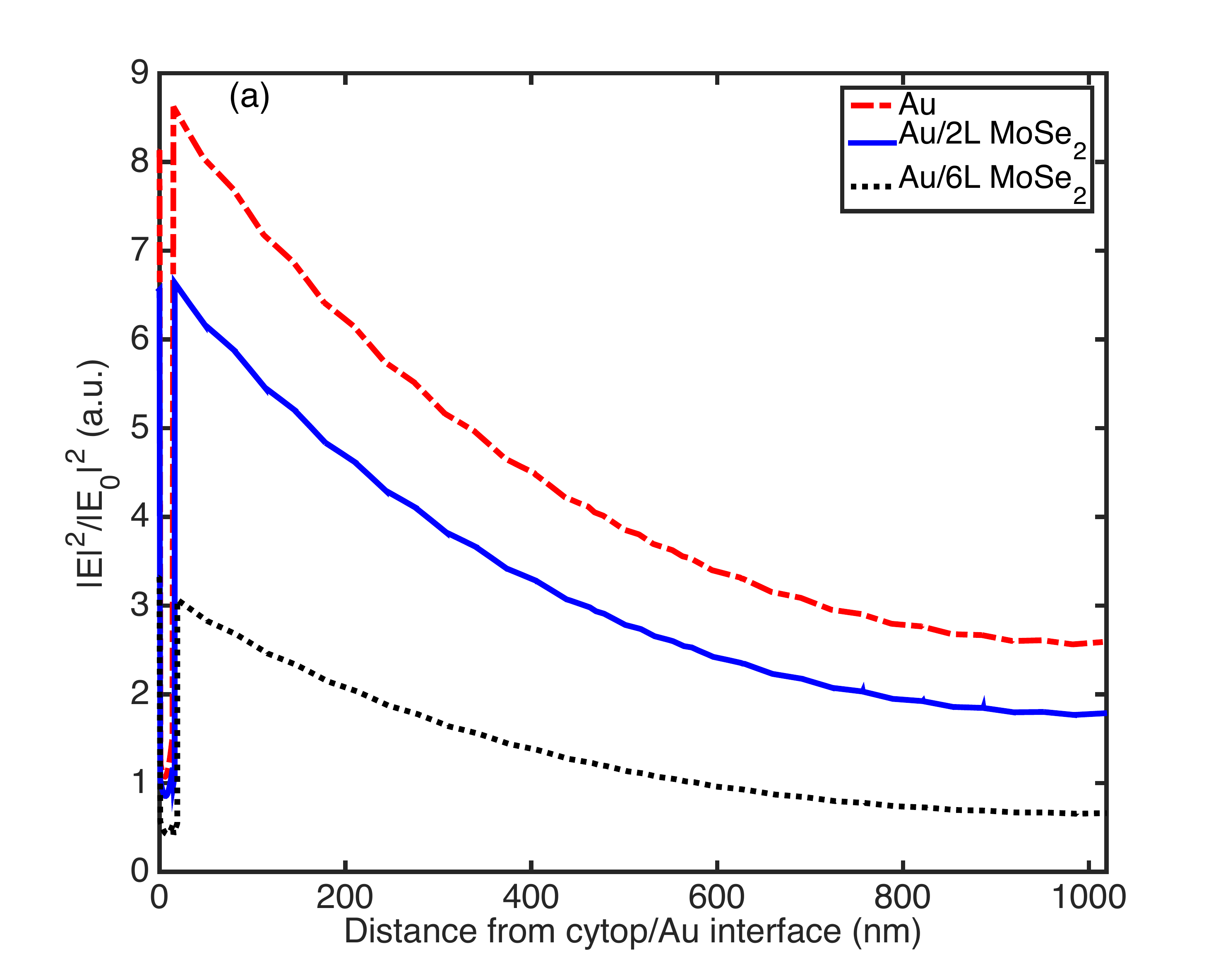}
      \includegraphics[scale=0.27]{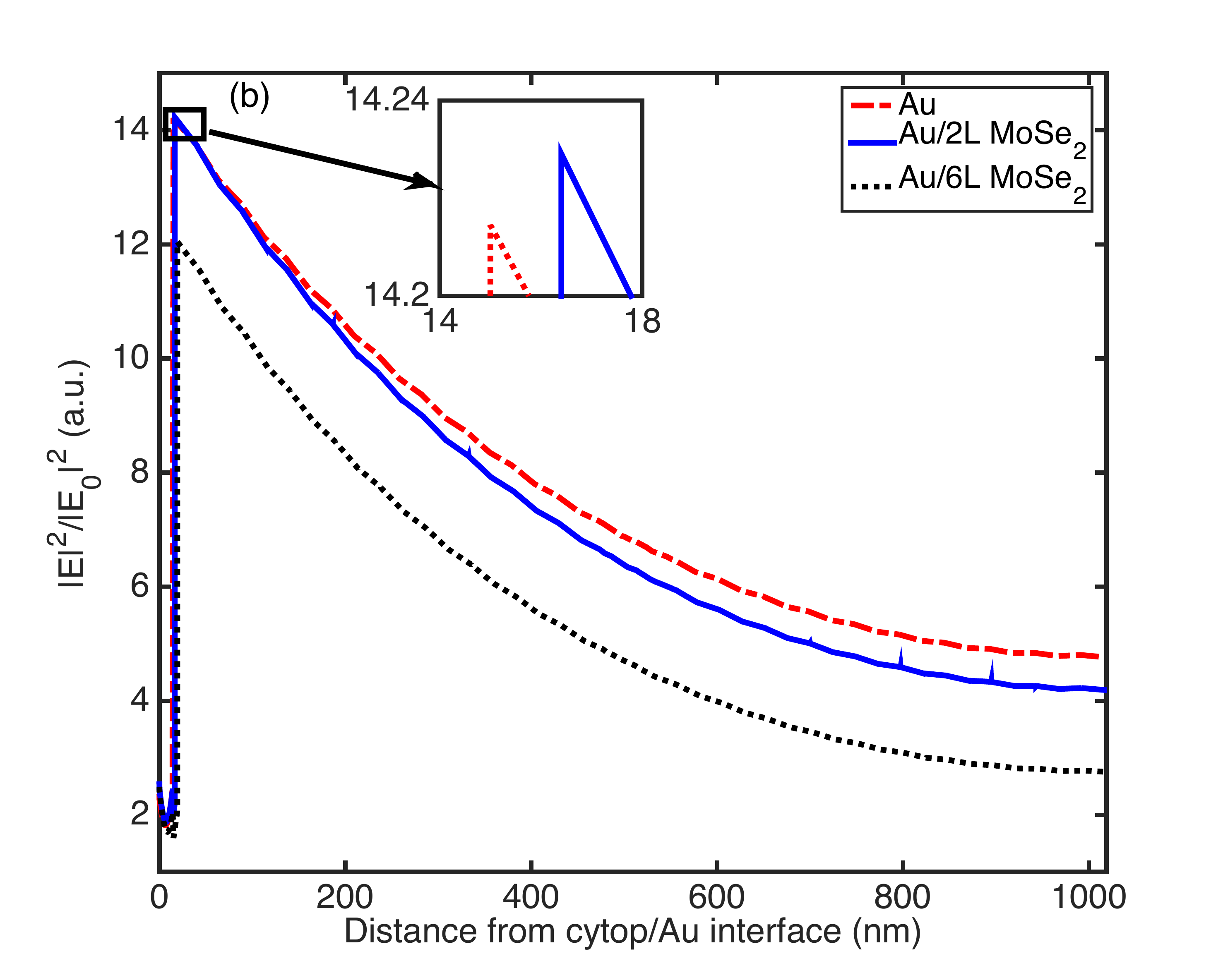}        
      \caption{ Electric field distribution for LRSPR sensors without $\text{MoSe}_2$, with bilayer and six layer $\text{MoSe}_2$. The thickness of Au film is 15 nm and the cytop layer thickness is (a) 2.0 $\mu$m and (b) 1.0 $\mu$m.}
\label{fig8}
\end{figure}
Although the imaging sensitivity for LRSPR sensor with bare and monolayer TMDC coated Au film in case 1 is higher than that in case 2, the improved imaging sensitivity in case 2 becomes higher with two and more layers $\text{MoS}_2$ and $\text{MoSe}_2$ coated Au film, or with four and more layers of $\text{WS}_2$ and $\text{WSe}_2$ applied. In order to understand the enhancement and degradation of imaging sensitivity for LRSPR sensor, the electric field distributions of proposed LRSPR biosensors at their resonance angles without $\text{MoSe}_2$ layers, with bilayer and six layers of $\text{MoSe}_2$ are obtained using finite element method (COMSOL Multiphysics), as shown in Fig. \ref{fig8}. For case 1, bare Au-based LRSPR sensor exhibits the highest electric field at the Au/sensing layer interface. The deposition of $\text{MoSe}_2$ decreases the intensity of evanescent electric field at the $\text{MoSe}_2$/sensing layer interface, which results in a degradation of imaging sensitivity. In case 2, coating Au film with bilayer $\text{MoSe}_2$ enhances the electric field norm at the $\text{MoSe}_2$/sensing layer interface as compared to that at Au film/sensing layer interface of bare Au-based LRSPR sensor (Fig. \ref{fig7}(b)). However, the deposition of 6 layers of $\text{MoSe}_2$ causes a reduction of the electric field norm at $\text{MoSe}_2$/sensing layer interface which in turn degrades the imaging sensitivity of LRSPR sensor. 
\begin{figure}[thpb]
      \centering
      \includegraphics[scale=0.24]{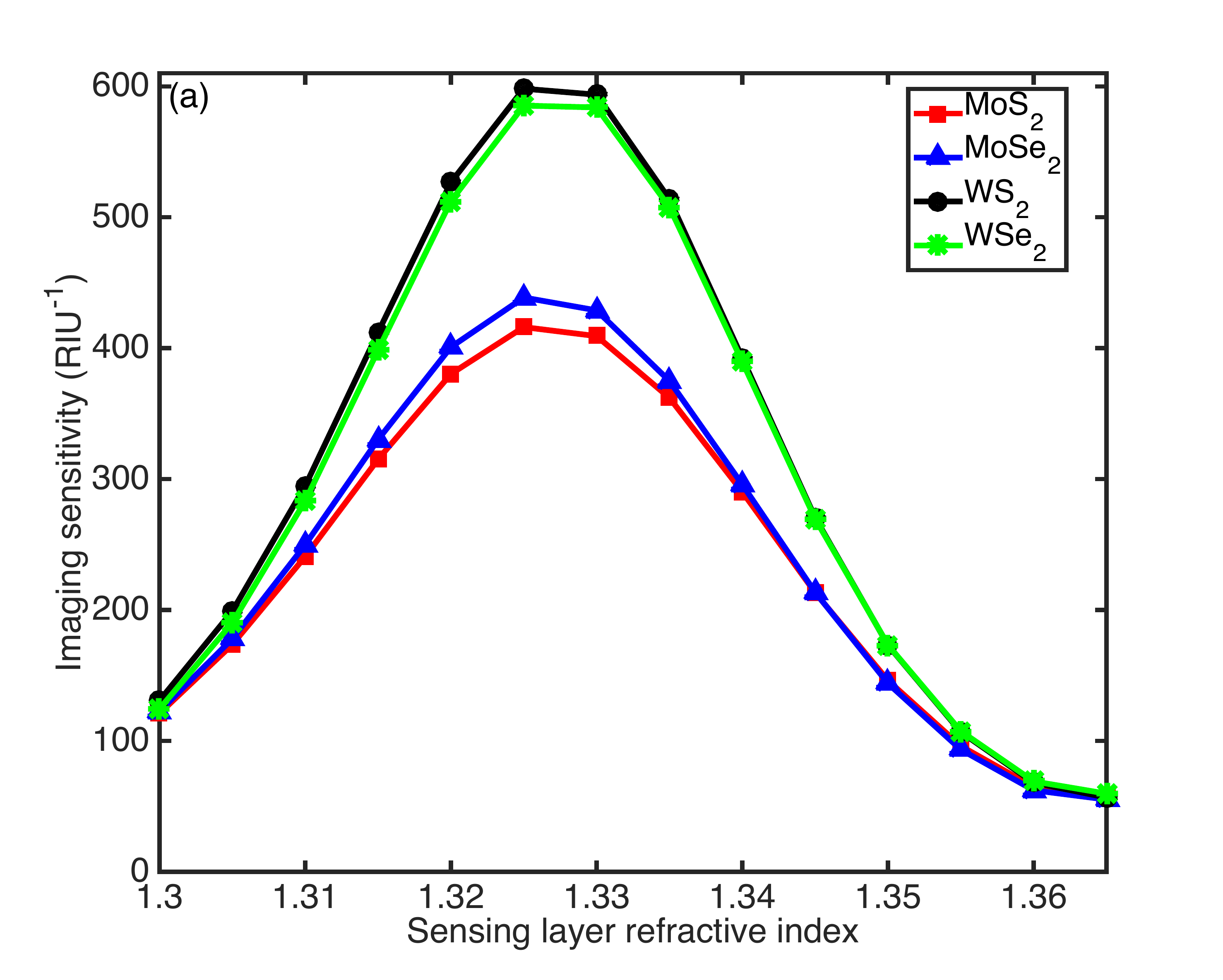}
      \includegraphics[scale=0.24]{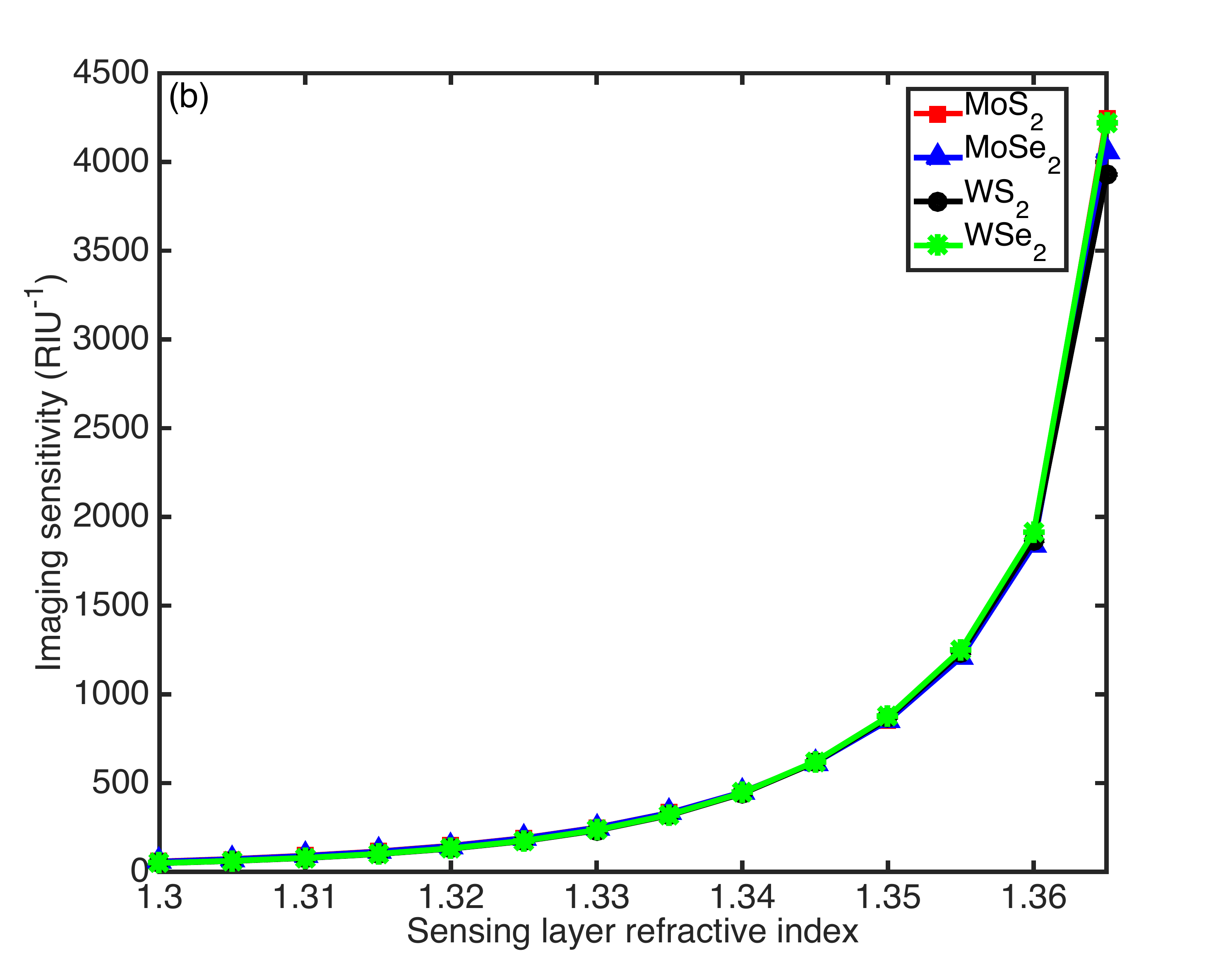}  
      \includegraphics[scale=0.24]{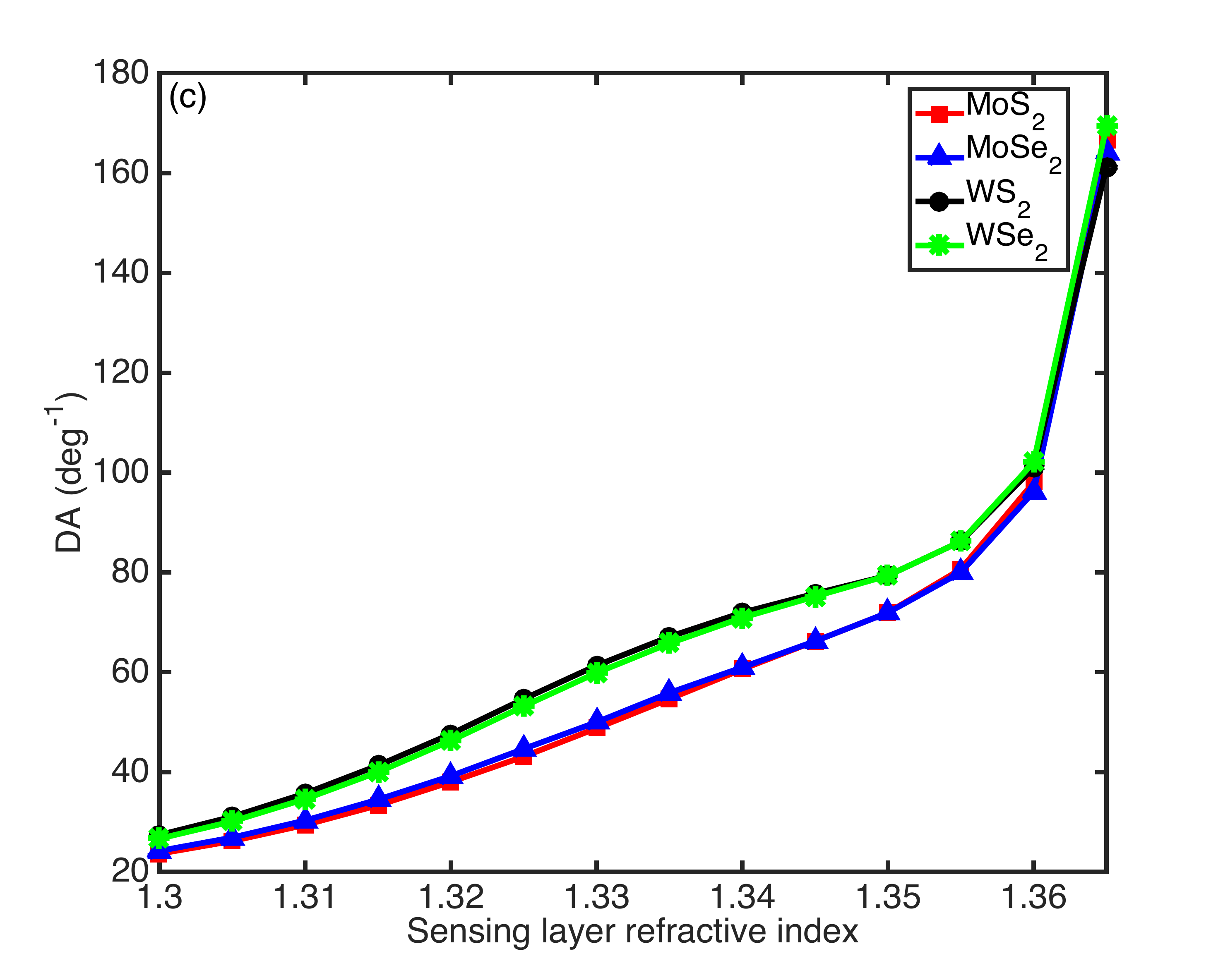}
      \includegraphics[scale=0.24]{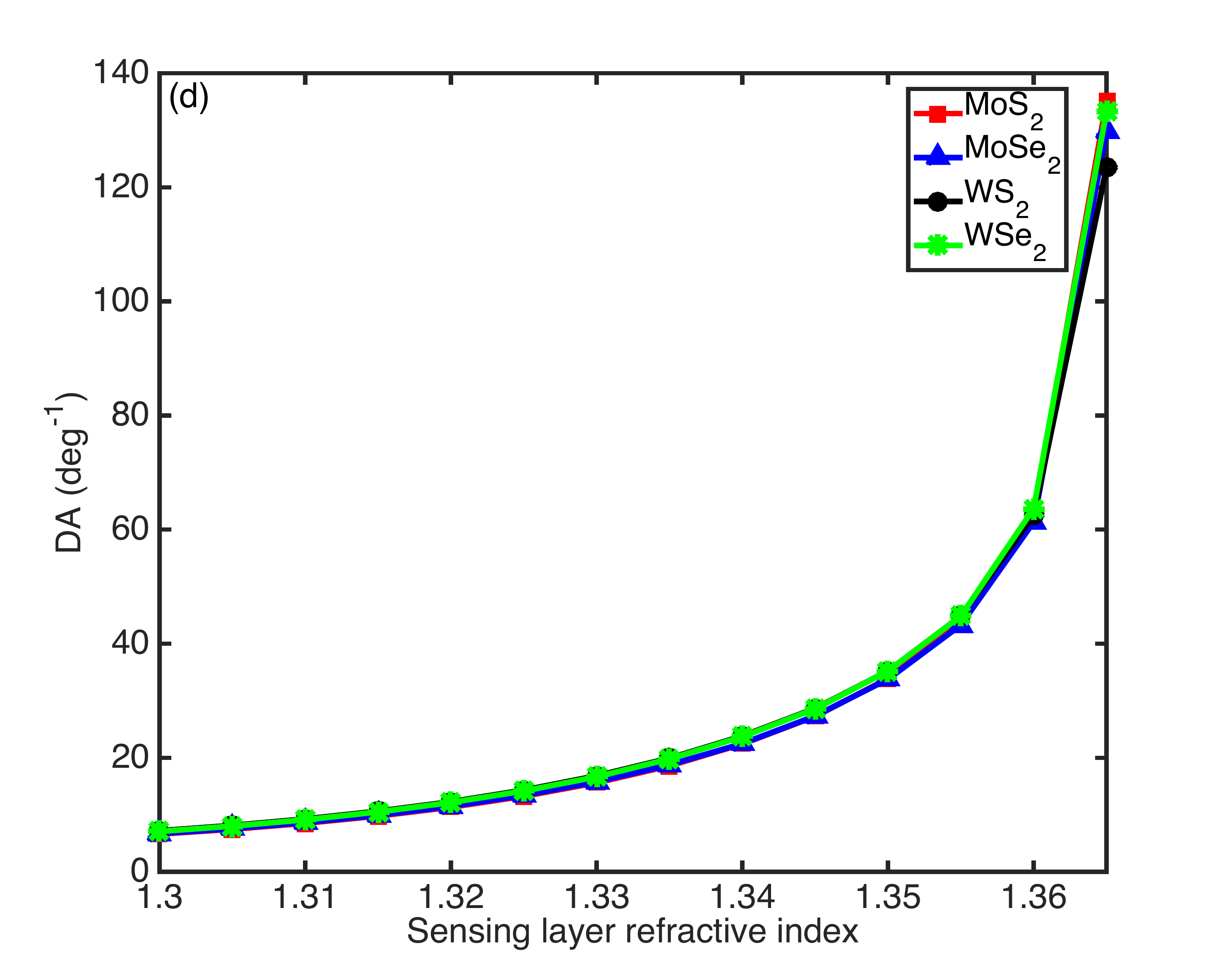}  
      \includegraphics[scale=0.24]{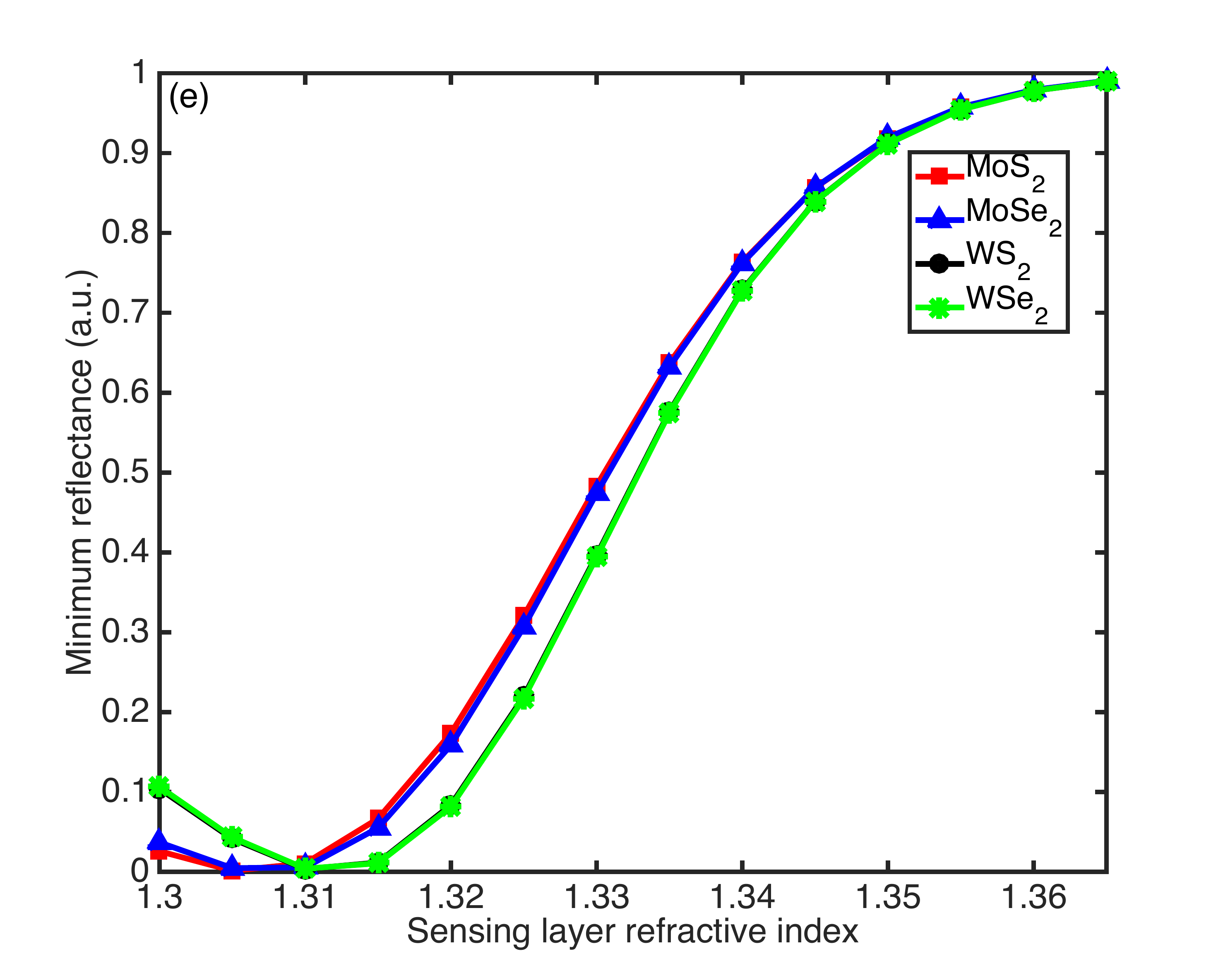}
      \includegraphics[scale=0.24]{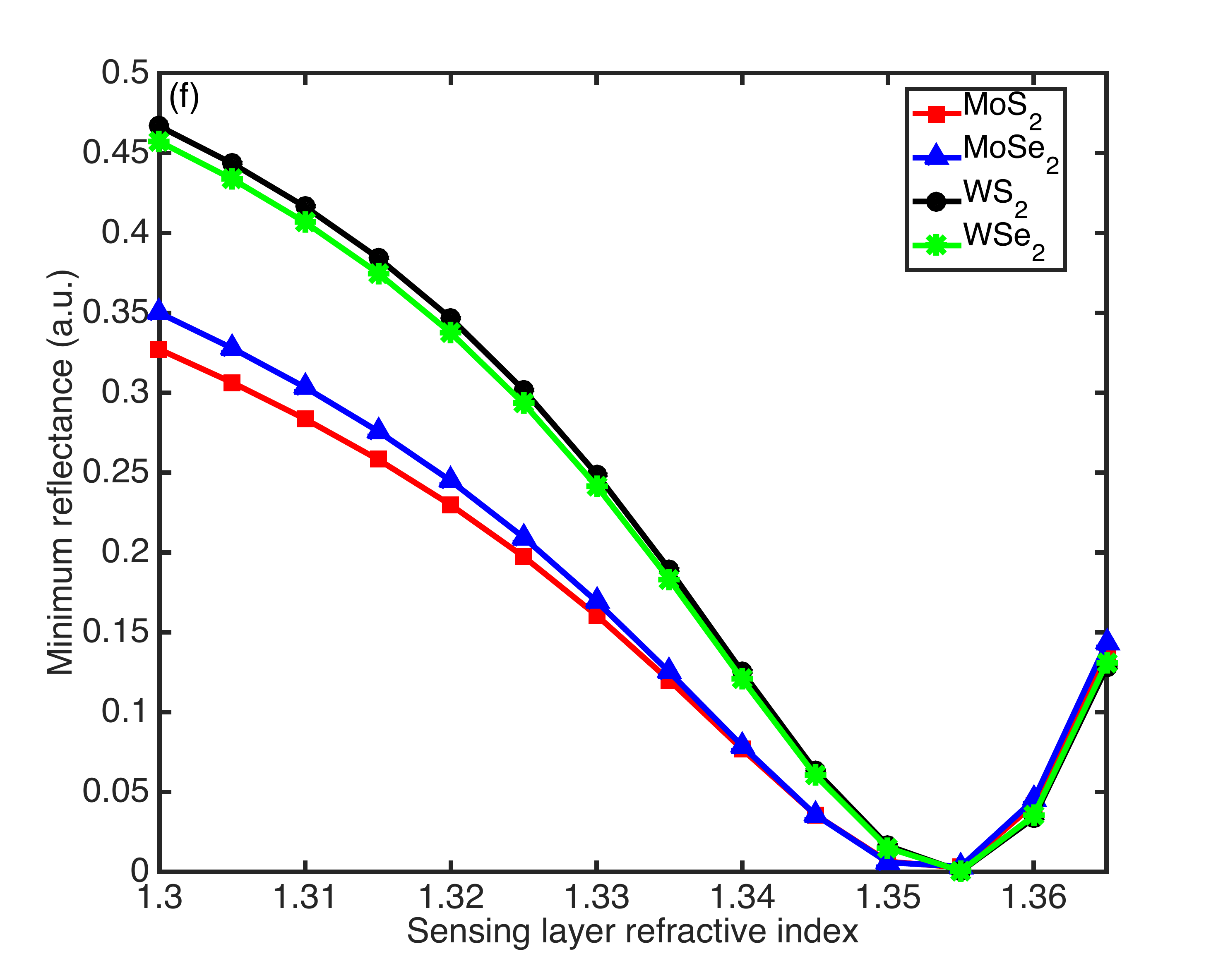}         
      \caption{Variation of (a)(b) imaging sensitivity, (c)(d) DA and (e)(f) minimum reflectance as a function sensing layer refractive index for monolayer TMDC-based LRSPR sensor. Au thin film thickness is 15 nm, and the thickness of cytop layer is (a)(c)(e) 2.0 $\mu$m and (b)(d)(f) 1.0 $\mu$m.}
\label{fig9}
\end{figure}

The RI of sensing layer is another important parameter for the sensor performance. The effect of sensing layer RI on the imaging sensitivity, DA and minimum reflectance is shown in Fig. \ref{fig9}. In case 1, the imaging sensitivity for the proposed LRSPR sensor firstly increased and then decreased in the RI range of 1.300-1.365. All the sensors get highest imaging sensitivity at RI of 1.325: 598.2 $\text{RIU}^{-1}$, 585.3 $\text{RIU}^{-1}$, 438.7 $\text{RIU}^{-1}$ and 416.1 $\text{RIU}^{-1}$ for monolayer $\text{WS}_2$-, $\text{WSe}_2$-, $\text{MoSe}_2$- and $\text{MoS}_2$-based sensor. The DA (Fig. \ref{fig9}(c)) increases with the sensing layer RI, and a high DA of more than 160 $\text{deg}^{-1}$ can be obtained at sensing layer RI of 1.365 for the proposed TMDCs-based LRSPR sensor. However, it should be noted that the minimum reflectance of the reflectance-angle curve is also high ($>0.9$) with sensing layer RI $\geq$1.350, and even higher than 0.99 at RI of 1.365 (see Fig. \ref{fig9}(e)), which limits the practical applications of the LRSPR sensor. Unlike case 1, the imaging sensitivity for LRSPR sensor in case 2 increases with the sensing layer RI (1.300-1.365), as shown in Fig. \ref{fig9}(b). The highest imaging sensitivity of 4244 $\text{RIU}^{-1}$, 4223 $\text{RIU}^{-1}$, 4060 $\text{RIU}^{-1}$ and 3926 $\text{RIU}^{-1}$ was obtained at RI of 1.365 for LRSPR sensor with monolayer $\text{MoS}_2$, $\text{WSe}_2$, $\text{MoSe}_2$, and $\text{WS}_2$, respectively. The DA in case 2 also increases with RI of sensing layer (see Fig. \ref{fig9}(d)), and the DA for $\text{MoS}_2$-, $\text{WSe}_2$-, $\text{MoSe}_2$- and , $\text{WS}_2$-based LRSPR sensor with ambient RI of 1.365 are 135.1 $\text{deg}^{-1}$, 133.3 $\text{deg}^{-1}$, 129.9 $\text{deg}^{-1}$ and 123.5 $\text{deg}^{-1}$, respectively. In addition, it can be seen from Fig. \ref{fig9}(f) that the minimum reflectance of reflectance-angle curve at higher RI of sensing layer ($>1.330$) is much smaller than that of LRSPR sensor in case 1.

It should be noted that the LRSPR sensor has a RI detection limit ($n_s=1.370$) in both cases. For sensing layer RI beyond this detection limit (i.e., $n_s\geq1.370$), the proposed LRSPR sensor can not detect the ambient RI variations. To improve the RI detection limit, the cytop layer with RI of 1.3395 should be replaced with other higher RI dielectric layers, such as magnesium fluoride ($\text{MgF}_2$) which has a RI of $\sim$1.38. Moreover, the sensor performance of the TMDCs-Au based LRSPR biosensors can be further improved by replace Au with other metals, such as silver, aluminium and copper, which has narrower SPR curve as compared to Au-based SPR sensor.  
\section{conclusion}
In this study, we explored an ultrasensitive LRSPR imaging biosensor with higher DA than that of cSPR biosensor, which in turn helps accurately measure the resonance dip. The proposed LRSPR biosensor consists of 2S2G prism, cytop layer, Au thin film, 2D TMDC ($\text{MoS}_2$/$\text{MoSe}_2$/$\text{WS}_2$/$\text{WSe}_2$), in which the TMDC layer serve as a signal-enhanced layer due to a high electron transfer efficiency from TMDC layer to Au surface, as well as a biomolecules absorption medium. In general, biosensors with Tungsten-based TMDCs ($\text{WS}_2$ and $\text{WSe}_2$) exhibit higher imaging sensitivity and DA than that of LRSPR biosensors with Molybdenum-based TMDCs ($\text{MoS}_2$ and $\text{MoSe}_2$). The sensitivity enhancement and degradation effect of the TMDCs-based LRSPR imaging biosensor that depends on the thickness of Au film and cytop layer has been investigated. An ultrahigh imaging sensitivity of 4244 $\text{RIU}^{-1}$, 4223 $\text{RIU}^{-1}$, 4060 $\text{RIU}^{-1}$ and 3926 $\text{RIU}^{-1}$ has been obtained at sensing layer RI of 1.365 for monolayer $\text{MoS}_2$-, $\text{WSe}_2$-, $\text{MoSe}_2$- and $\text{WS}_2$-based LRSPR biosensor, respectively, with a high DA of 135.1 $\text{deg}^{-1}$, 133.3 $\text{deg}^{-1}$, 129.9 $\text{deg}^{-1}$ and 123.5 $\text{deg}^{-1}$. The RI detection limit can be improved by replacing the cytop layer with another higher RI dielectric layer. In addition, the imaging sensitivity and DA of the proposed LRSPR biosensor can be further enhanced by using other metallic thin film, like silver, aluminium and copper. The proposed LRSPR imaging sensors based on TMDCs are potentially useful in chemical and biosensing applications for simultaneous detection of multiple biomolecular interactions.

\begin{acknowledgments}
This work was supported by Singapore ASTAR AME IRG A1783c0011. 
\end{acknowledgments}



%

\end{document}